\documentclass[a4paper,12pt]{article}

\usepackage{amssymb,amsmath}
\usepackage{latexsym}
\usepackage{graphicx}
\usepackage[usenames,dvipsnames]{xcolor}
\usepackage{psfrag}
\usepackage[pdftex,colorlinks]{hyperref}
\hypersetup{colorlinks=true,linkcolor=blue,urlcolor=blue,filecolor=blue,citecolor=blue}
\usepackage{cite}
\usepackage{mathrsfs}
\usepackage{fancyhdr}
\usepackage{enumerate}
\usepackage{bm}

\usepackage{empheq} \empheqset{box=\fbox}

\usepackage{shorttoc} 
\usepackage{multicol}

\usepackage{pgfplots}
\usetikzlibrary{decorations.markings,decorations.pathmorphing,patterns,calc,snakes,arrows}

\makeatletter

\def\AdSxS{\mbox{${\rm AdS}_3\times S^3$}}

\definecolor{DarkGreen}{rgb}{0,.5,0}%
\definecolor{lightgray}{rgb}{.8,.8,.8}%
\definecolor{darkgray}{rgb}{.3,.3,.3}%

\g@addto@macro\bfseries{\boldmath}

\def\refchecklabelfontsize{\fontsize{5pt}{5pt}\selectfont}
\let\mark@size=\refchecklabelfontsize

\def\half{{\frac{1}{2}}}

\def\p{\partial}

\def\unit{{1\kern-.65ex {\rm l}}}
\def\1{{1\kern-.65ex {\rm l}}}

\def\Re{\mathop{\mathrm{Re}}\nolimits}

\def\ket#1{{|{#1}\rangle}}

\def\kett#1{{|{#1}\rangle\!\rangle}}

\def\rt{{\widetilde{r}}}

\def\yt{{\widetilde{y}}}

\def\Gt{{\widetilde{G}}}

\def\Jt{{\widetilde{J}}}

\def\Lt{{\widetilde{L}}}

\def\Tt{{\widetilde{T}}}

\def\etat{{\widetilde{\eta}}}

\def\phit{{\widetilde{\phi}}}
\def\psit{{\widetilde{\psi}}}

\def\hb{{\bar{h}}}
\def\ib{{\bar{\imath}}}
\def\jb{{\bar{\jmath}}}

\def\zb{{\bar{z}}}

\def\Gb{{\bar{G}}}
\def\Jb{{\bar{J}}}
\def\Lb{{\bar{L}}}

\def\Tb{{\bar{T}}}

\def\etab{{\bar{\eta}}}

\def\cB{{\cal B}}

\def\cD{{\cal D}}

\def\cF{{\cal F}}

\def\cM{{\cal M}}
\def\cN{{\cal N}}
\def\cO{{\cal O}}
\def\cP{{\cal P}}

\def\cR{{\cal R}}

\def\bbR{{\mathbb{R}}}

\def\bbZ{{\mathbb{Z}}}

\newcount\hour \newcount\minute
\hour=\time \divide \hour by 60
\minute=\time
\def\now{%
\ifnum \hour<13
  \ifnum \hour=0 \advance \hour by 12 \number\hour:\else \number\hour:\fi%
     \ifnum \minute<10 0\fi%
     \number\minute%
\ A.M.%
\else \advance \hour by -12 \number\hour:%
  \ifnum \minute<10 0\fi%
  \number\minute%
  \ P.M.%
\fi%
}

\makeatother

\begin{document}

\baselineskip=18pt  
\numberwithin{equation}{section}  







\thispagestyle{empty}

\vspace*{-2cm} 
\begin{flushright}
YITP-22-162\\
\end{flushright}

\vspace*{2.5cm} 
\begin{center}
 {\LARGE Superstrata on Orbifolded Backgrounds}\\
 \vspace*{1.7cm}
 Masaki Shigemori\\
 \vspace*{1.0cm} 
Department of Physics, Nagoya University\\
Furo-cho, Chikusa-ku, Nagoya 464-8602, Japan\\
and\\
Center for Gravitational Physics,\\
Yukawa Institute for Theoretical Physics, Kyoto University\\
Kitashirakawa Oiwakecho, Sakyo-ku, Kyoto 606-8502, Japan
\end{center}
\vspace*{1.5cm}

\noindent
Some microstates of the Strominger-Vafa black hole are represented by
smooth horizonless geometries called superstrata.  The standard
superstrata are deformations of ${\rm AdS}_3\times S^3$, but there are
also generalizations of superstrata on the orbifold $({\rm AdS}_3\times
S^3)/\mathbb{Z}_p$.  In this paper, we discuss aspects of such
orbifolded superstrata.  We present a CFT perspective on the structure
of orbifolded superstrata, showing that they can be constructed in a
$p$-covering space of the orbifold CFT just as the standard superstrata.
We also explicitly write down and study the geometry of the orbifolded
superstrata, focusing on the difference from the non-orbifold case,
$p=1$.  We discuss further generalization of superstrata to the ones on
a fractional spectral flow of $({\rm AdS}_3\times S^3)/\mathbb{Z}_p$.
This generalization involves new fractional mode excitations of the CFT
side.  We estimate the number of those generalized superstrata, and show
that their entropy is too small to account for the Strominger-Vafa
entropy.  We will discuss some implications of this result, related to
the typical microstates of the black hole and the relevant supersymmetry
index.

\newpage
\setcounter{page}{1} 






\section{Introduction}

It has been known for a while that some microstates of black holes in
string theory are represented by smooth, horizonless geometries in
classical gravity called microstate geometries \cite{Bena:2013dka,
Bena:2022ldq, Bena:2022rna}.  An arena for the study of microstate
geometries is the D1-D5 system -- type IIB superstring in $\bbR_t\times
\bbR^4\times S^1\times \cM$ where $\cM=T^4$ or $\rm K3$, with $N_1$
D1-branes wrapping $S^1$ and $N_5$ D5-branes wrapping $S^1\times \cM$.
The fact that this system has $\AdSxS\times \cM$ near-horizon geometry
and thus allows a holographic CFT description via the ${\rm
AdS_3/CFT_2}$ correspondence is quite useful in analyzing various
physical properties of microstate geometries.
By adding to the D1-D5 system $N_P$ units of momentum along $S^1$, we
can construct a 3-charge D1-D5-P black hole with area entropy
\begin{align}
 S_{\rm BH}=2\pi\sqrt{N N_P},\qquad N\equiv N_1 N_5\label{loyl20Dec22}
\end{align}
which was reproduced by Strominger and Vafa \cite{Strominger:1996sh} by
counting microstates in the brane worldvolume theory.  Many microstate
geometries for this black hole have been constructed, giving a bulk
gravity picture of these microstates.  However, they are atypical states
of the black-hole ensemble in that their entropy is parametrically
smaller than \eqref{loyl20Dec22}.

The microstate geometries constructed so far roughly fall into two
classes. The first one is called multi-center bubbled solutions which
can be regarded as bound states of D-branes that have transitioned into
fluxes \cite{Bena:2004de, Gauntlett:2004qy, 
Bena:2005va, Berglund:2005vb,
Bena:2006kb, Bena:2007kg,
Bena:2007qc, Bena:2010gg, Bianchi:2017bxl, Heidmann:2017cxt, Bena:2017fvm,
Avila:2017pwi, Mayerson:2022yoc, Rawash:2022sum}.  
The CFT duals of
multi-center solutions are generally unknown, except for two-center
cases \cite{Giusto:2004id, Giusto:2004ip, Lunin:2004uu, Jejjala:2005yu,
Giusto:2012yz, Chakrabarty:2015foa}.
The second class is called superstrata \cite{Bena:2015bea, Bena:2016agb,
Bena:2016ypk, Bena:2017geu, Bena:2017xbt, Ceplak:2018pws,
Heidmann:2019zws, Heidmann:2019xrd, Ganchev:2022exf, Ceplak:2022pep,
Shigemori:2020yuo} that represent a coherent, backreacted gas of
supergravitons inside $\AdSxS$.  The CFT duals of superstrata are
well understood because they are clear from the construction.
To begin with, the dictionary between single-particle 1/4-BPS
supergraviton states\footnote{\label{ftnt:convention_BPSness}This
preserves half (eight) of the 16 real supercharges of the $\AdSxS$
background, but we use the convention in which the number of real
supercharges is counted relative to the 32 supersymmetries of the
maximal supergravity.  So, 1/4-BPS means eight supersymmetries, and
1/8-BPS means four.} in the bulk and single-particle chiral primaries in
CFT is well-known \cite{Maldacena:1998bw, Deger:1998nm, Larsen:1998xm,
deBoer:1998kjm}.  In the bulk, by acting with Killing vectors of
$\AdSxS$ on a single-particle state, we can generate single-particle
1/8-BPS states.  Coherent multi-particle excitations of such 1/8-BPS
supergravitons are nothing other than superstrata.
On the CFT side, we can act with Virasoro and current algebra generators
on chiral primaries to generate descendants that are dual to 1/8-BPS
supergravitons. By tensoring such descendants, we get states dual to the
bulk superstrata.  So, the constructions are parallel in CFT and in the
bulk, suggesting a map between CFT and superstratum states, as has been
confirmed by explicit computations \cite{Giusto:2015dfa, Giusto:2019qig,
Rawash:2021pik, Ganchev:2021ewa} (but see also \cite{Guo:2022ifr}).

Superstrata are the most general microstate geometries with known CFT
duals. As a non-trivial but tractable playground, their physical
properties and implications for black-hole physics have been very
actively investigated \cite{Tyukov:2017uig, Bena:2018mpb, Raju:2018xue,
Bena:2018bbd, Bianchi:2018kzy, Bena:2019azk, Bombini:2019vnc,
Tian:2019ash, Bena:2020iyw, Martinec:2020cml, Ceplak:2021kgl,
Mayerson:2020tpn, Bacchini:2021fig, Bah:2021jno, Ikeda:2021uvc}.
However, as already mentioned, their entropy is far smaller than the
Strominger-Vafa entropy \eqref{loyl20Dec22} \cite{Shigemori:2019orj,
Mayerson:2020acj}.  In CFT terminology, superstratum states are
constructed based on rigid generators of the Virasoro and current
algebras, such as $L_{-1}$.  To reproduce the Strominger-Vafa entropy,
more general, fractional and higher modes are needed, such as
$L_{-{1\over N}}$ or $J^+_{-5}$.

In \cite{Bena:2016agb}, superstrata with some special fractional modes
were constructed and, later, in~\cite{Shigemori:2020yuo}, more general
cases with fractional modes were discussed.  These are superstrata on
the orbifolds of $\AdSxS$.  The idea is fairly simple; global $\AdSxS$
of the D1-D5 system can be quotiented by $\bbZ_p$ by making the radius
$R_y$ of the $S^1$ circle direction $p$ times smaller, which can be
effected by replacing $R_y\to p R_y$ in the solution.  This procedure
gives an orbifold that we call $\AdSxS/\bbZ_p$.\footnote{It is more
accurate to call it $(\AdSxS)/\bbZ_p$, but for notational simplicity we
denote it by $\AdSxS/\bbZ_p$.}  The $\bbZ_p$ orbifold singularity
corresponds to that of $p$ coincident Kaluza-Klein monopoles, which is
allowed in string theory.  When we have a superstratum on $\AdSxS$, we
cannot in general quotient the solution by $\bbZ_p$, because the $S^1$
direction is not an isometry; the superstratum solution involves a
propagating wave along it.  However, when the wave number satisfies a
quantization condition, the solution after division by $\bbZ_p$ remains
single-valued and represents a valid solution.

In this paper, we will discuss various aspects of 
superstrata on orbifolded $\AdSxS$ backgrounds.
First, we will explain the construction in CFT, showing that the
procedure of constructing superstrata -- starting with chiral primaries and
acting on it with symmetry generators, which worked for the non-orbifold
case -- can be extended to the orbifold case.  More specifically, by going
to a covering space in the D1-D5 symmetric orbifold CFT, we can consider a
covering-space version of chiral primaries.  By acting on them with a
covering-space version of symmetry generators, which are fractional
generators in the base space, we can generate descendants.  Coherent
superpositions of those generalized descendants are the superstrata on
$\AdSxS/\bbZ_p$.  This procedure is completely in parallel with the bulk
where we construct superstrata in $\AdSxS$, which is the covering space,
and then divide the solution by $\bbZ_p$ to go to the ``base space''
$\AdSxS/\bbZ_p$.  In both CFT and the bulk, the generators in the covering
space become fractional generators in the real, base space, such as
$L_{-{1\over p}},J^+_{-{1\over p}}$.

We write down the explicit form of some superstrata on the orbifold
$\AdSxS/\bbZ_p$ and analyze their geometries.  As an example, we will
study the orbifold version of the so-called $(1,0,n)$ superstratum,
which we call $(1,0,n)_p$.  This has a long throat that starts deeper
inside the geometry than the non-orbifold version.  The throat is
narrower than the non-orbifold version but its length is the same.  At
the bottom of the throat, the geometry ends with a $\bbZ_p$ conical
singularity.  Because the form of the $(1,0,n)_p$ geometry is as simple
as $(1,0,n)$, it would be interesting to study its physical properties,
focusing on the effect of $p>1$ as compared to the $p=1$ case
\cite{Tyukov:2017uig, Bena:2018mpb, Bena:2019azk, Bena:2020iyw, Martinec:2020cml, Ceplak:2021kgl}.

Furthermore, we generalize superstrata on the $\AdSxS/\bbZ_p$ background
to superstrata on the fractional spectral flow of $\AdSxS/\bbZ_p$, which
we call $(\AdSxS/\bbZ_p)_{[{s\over p}]}$, $s\in\bbZ$.  These superstrata
now involve more general fractional modes such as $J^{+}_{-{1+2s\over
p}}$.  Actually, this background $(\AdSxS/\bbZ_p)_{[{s\over p}]}$ is
nothing but the so-called GLMT geometry \cite{Giusto:2012yz}, the most
general 2-center bubbled geometry.  The original superstrata are
coherent supergravitons on a very special 2-center bubbled geometry, but
superstrata on $(\AdSxS/\bbZ_p)_{[{s\over p}]}$ are coherent
supergravitons on the most general 2-center bubbled geometry.

Because we thus have generalized superstrata that involve various
fractional modes, it is natural to ask if their entropy reproduces the
Strominger-Vafa entropy.  We will estimate the entropy of superstrata on
$\AdSxS/\bbZ_p$, and find that its entropy is actually even smaller than
the entropy of the original superstrata.  The fractional modes that
those generalized superstrata contain are very special ones and do not
lead to enhancement of entropy.  However, they are expected to
contribute to supersymmetry indices. We will discuss some of its
interesting implications.

The structure of the rest of the present paper is as follows.  In
section \ref{s:strata_in_D1-D5_CFT}, we will discuss subjects that will
be the basis for the discussions in the subsequent sections.  We will
briefly review how superstrata around $\AdSxS$ are constructed, in both
the D1-D5 CFT and in the dual supergravity.  We will also discuss
aspects of the orbifold background $\AdSxS/\bbZ_p$, and of the spectral
flow transformation in CFT and in the bulk.
In section \ref{sec:CFT_view}, we will discuss the CFT perspective on
the construction of superstrata on $\AdSxS/\bbZ_p$.  Focusing on CFT
states made of strands whose lengths are integer multiples of $p$, we
will discuss how to go to a $p$-covering space and define currents
defined on the $p$-cover.  Then, we will show that anti-chiral primaries
in the base space lift to anti-chiral primaries in the covering space
and, based on them, we can generate states that can be interpreted as
superstrata on $\AdSxS/\bbZ_p$. 
In section \ref{sec:gravity_side}, we will study the gravity aspects of
superstrata on $\AdSxS/\bbZ_p$. We will take some specific examples and
study their geometry, contrasting them with the non-orbifold $p=1$ case.
In section \ref{sec:fract_spectral_flow}, we will give a further
generalization of superstrata.  we will consider superstrata on what we
call $(\AdSxS/\bbZ_p)_{[{s\over p}]}$, which is a fractional spectral
flow of $\AdSxS/\bbZ_p$.  In CFT, these superstrata contains new
fractional modes such as $J^+_{-{1+2s\over p}}$.
In section \ref{sec:disc}, we will discuss some aspects of the
orbifolded superstrata.  We will estimate their entropy and find that it
is even smaller than that of the ordinary superstrata.  Namely, they are
too few to account for the Strominger-Vafa entropy.  We will discuss
some implications of this fact.
In Appendix \ref{app:sugra_soln}, we will summarize supergravity
solutions that we use in the main text.

\section{Preliminaries}

\label{s:strata_in_D1-D5_CFT}

\subsection{D1-D5 CFT}
\label{ss:D1-D5_CFT}

We briefly recall the properties of the D1-D5 CFT that are relevant to
the current paper.\footnote{For detail of the D1-D5 CFT, see
e.g.~\cite{David:2002wn, Avery:2010qw}.  For its aspects relevant to
superstrata, see e.g.~\cite{Shigemori:2020yuo, Bena:2017xbt}.}
This theory is a $d=2,\cN=(4,4)$ CFT with a symmetry group
$SU(1,1|2)_L\times SU(1,1|2)_R$, with the associated currents
$T(z),G^{\alpha A}(z),J^i(z)$ and modes $L_n,G^{\alpha
A}_n,J^i_n$.  The right-moving versions of the currents are
$\Tb(\zb),\Gb^{\dot\alpha A}(\zb),\Jb^\ib(\zb)$ with modes
$\Lb_n,\Gb^{\dot{\alpha} A}_n,\Jb^{\,\ib}_n$.  Here, $\alpha=\pm$
($i=1,2,3$) is a doublet (triplet) index for an R-symmetry group
$SU(2)_L\subset SU(1,1|2)_L$, and $\dot{\alpha},\ib$ are their
right-moving counterparts.  At the so-called orbifold point in its moduli
space, this theory is described by a symmetric orbifold CFT with target
space ${\rm Sym}^N \cM$ with $\cM=T^4$ or K3.  Here we focus on
$\cM=T^4$, for which $N=N_1 N_5$.  The theory also has a global
$SU(2)\times SU(2)$ symmetry whose doublet indices are denoted by
$A,\dot{A}$, although this symmetry is broken by the compactness of
$\cM$.

As a theory with $\cN \ge 2$ supersymmetry, the D1-D5 CFT in the NS
sector has chiral primaries which are 1/4-BPS states\footnote{See
footnote \ref{ftnt:convention_BPSness}.}.  Actually, for convenience in
matching with the bulk convention, we consider anti-chiral primaries. A
left--anti-chiral primary $\ket{\psi}$ is annihilated not only by all
positive-mode generators, $L_{n},J^i_{n},G_{n-\half}^{\alpha A}$ with
$n\ge 1$, but also by $G_{-\half}^{-A},J^-_0$.  Or equivalently, it
satisfies $h=-j$ where $(h,j)$ are the eigenvalues of ($L_0,J^3_0$) with
$h\ge 0$.  A full anti-chiral primary also satisfies $\hb=-\jb$ for the
right-moving sector.

The general anti-chiral primaries of the D1-D5 CFT are constructed by
tensoring the following single-particle anti-chiral primaries:
\begin{align}
\begin{aligned}
 &\ket{\alpha\dot\alpha}_k,&\quad
 h&=-j=\tfrac{k-\alpha}{2},&
 \hb&=-\jb=\tfrac{k-\dot{\alpha}}{2},\\
 &\ket{\alpha \dot{A}}_k,&
 h&=-j=\tfrac{k-\alpha}{ 2},&
 \hb&=-\jb=\tfrac{k}{ 2},\\
 &\ket{\dot{A}\dot\alpha}_k,&
 h&=-j=\tfrac{k}{ 2},&
 \hb&=-\jb=\tfrac{k-\dot\alpha}{ 2},\\
 &\ket{\dot{A}\dot{B}}_k,&
 h&=-j=\tfrac{k}{ 2},& 
 \hb&=-\jb=\tfrac{k}{ 2},
\end{aligned}
\label{T4_1-part_ch_pr}
\end{align}
At the orbifold point, these states correspond to twist operators of
order $k$; namely, they glue $k$ copies of $\cM $ (out of $N$ copies).
These $k$ copies glued together are called a strand of length~$k$.  The
$SU(2)$ invariant combination $
\tfrac{1}{\sqrt{2}}\epsilon_{\dot{A}\dot{B}}\ket{\dot{A}\dot{B}}_k$ is
denoted by $\ket{00}_k$.
Among the states in \eqref{T4_1-part_ch_pr}, the state
$\ket{++}_1=\ket{\alpha=+,\dot\alpha=+}_1$ plays a special role because
it has $h=j=\hb=\jb=0$ and represents the NS vacuum (of a single copy
of $\cM $).

The general anti-chiral primaries are obtained by tensoring together
single-particle ones so that the total length is $N$ as follows:
\begin{align}
 \prod_{\psi}
 \prod_{k=1}^N
 \bigl[\ket{\psi}_k\bigr]^{N^{\psi}_k},
\quad\text{with}\quad
 \sum_\psi \sum_k k N^\psi_{k}=N.
\label{gen_ch_pr_cft}
\end{align}
where $\ket{\psi}$ runs over different species in
\eqref{T4_1-part_ch_pr}.  This state clearly satisfies $h=-j$.
For example, we can have a state like
\begin{align}
  \ket{00}_k ~ (\ket{--}_{k'})^3 ~ 
 \underbrace{\ket{++}_1\dots\ket{++}_1}_{\text{$N-k-3k'$ strands}}.
\label{exampleXp_k=1}
\end{align}
In the bulk, such states correspond to multi-particle, 1/4-BPS states of
supergravitons
\cite{Maldacena:1998bw, Deger:1998nm, Larsen:1998xm, deBoer:1998kjm}.  In~\eqref{exampleXp_k=1}, the trivial $\ket{++}_1$ part
represents the \AdSxS\ background, while $\ket{00}_k$ and
$\ket{--}_{k'}$ are supergravitons in the background.

\subsection{Descendants and superstrata}

By acting with the generators $\{L_{-1},G_{-\half}^{+A},J_0^+\}$ on a
single-particle anti-chiral primary, we can generate (super)descendant
states with respect to the rigid $SU(1,1|2)_L\times SU(1,1|2)_R$
symmetry.  Concretely, if we start with an anti-chiral primary with
$h=-j$ denoted by $\kett{j,-j}$, where $j\ge 0$, we generate the
following states:
\begin{subequations} 
 \label{members_of_short_multiplet}
 \begin{align}
 & \textstyle
 \kett{j+n,-j}\xrightarrow{J_0^+}\kett{j+n,-j+1}\xrightarrow{J_0^+}\cdots\xrightarrow{J_0^+}\kett{j+n,j}
 \label{members_of_short_multiplet1}
 \\
 &G_{-\half}^{+,A}\bigg\downarrow\notag\\
 &\textstyle
 \kett{j+\half+n,-j+\half}\xrightarrow{J_0^+}\kett{j+\half+n,-j+{3\over 2}}\xrightarrow{J_0^+}\cdots\xrightarrow{J_0^+}\kett{j+\half+n,j-\half}
 \label{members_of_short_multiplet2}
 \\
 &G_{-\half}^{+,B}\bigg\downarrow\notag\\
 &\textstyle
 \kett{j+1+n,-j+1}\xrightarrow{J_0^+}\kett{j+1+n,-j+2}\xrightarrow{J_0^+}\cdots\xrightarrow{J_0^+}\kett{j+1+n,j-1}
 \label{members_of_short_multiplet3}
 \end{align}
\end{subequations}
Here, $\kett{h,j}$ means a state with $(L_0,J_0^3)=(h,j)$.  The states
in the second line are doubly degenerate, because $A=1,2$.  The third
line has no such degeneracy because we can only descend from the first
line with $G_{-\half}^{+,1}G_{-\half}^{+,2}$.  More precisely, to get a
state orthogonal to other states, we must act instead with
$G_{-\half}^{+,1}G_{-\half}^{+,2}+{1\over 2h}L_{-1}J_0^+$ where $h$ is
the value of $L_0$ for the anti-chiral primary \cite{Avery:2010qw,
Ceplak:2018pws}.  Moreover, the number $n=0,1,\dots$ corresponds to the
number of times we act on the state with $L_{-1}$.  Explicitly, the
three lines in \eqref{members_of_short_multiplet} are the 
following states:
\begin{subequations} 
 \label{1p_1/8_sugrtn}
  \begin{align}
 (J_0^+)^m (L_{-1})^n& \ket{\psi}_k, \label{1p_1/8_sugrtn1}\\
 (J_0^+)^m (L_{-1})^n G_{-\half}^{+,A}& \ket{\psi}_k,\label{1p_1/8_sugrtn2}\\
 \hspace*{-5ex}
   (J_0^+)^m (L_{-1})^n \left(G_{-\half}^{+,1}G_{-\half}^{+,2}+\tfrac{1}{2h}L_{-1}J_0^+\right)
   &\ket{\psi}_k,\label{1p_1/8_sugrtn3}
 \end{align}
\end{subequations}
where $m\ge 0$ and its maximum possible value is found from
\eqref{members_of_short_multiplet}.
The states \eqref{1p_1/8_sugrtn} are generally 1/8-BPS,\footnote{See footnote \ref{ftnt:convention_BPSness}. } breaking all
left-moving supersymmetry.  In the bulk, they correspond to
single-particle, 1/8-BPS supergraviton states.

Just as in the 1/4-BPS case, we can tensor together single-particle
1/8-BPS states to construct a more general, multi-particle 1/8-BPS
state.  For example, we can have
\begin{align}
 \left[\ket{++}_1\right]^{N_0} \prod_{k,m,n,q} 
 \left[(J_0^+)^m(L_{-1})^n 
\left(G_{-\half}^{+,1}G_{-\half}^{+,2}+\tfrac{1}{2h}L_{-1}J_0^+\right)^q
 \ket{00}_k\right]^{N_{kmnq}}
 \label{mebn31Oct22}
\end{align}
with
\begin{align}
 N_0+\sum_{k,m,n,q} k N_{kmnq}=N,\qquad
 q=0,1.
\end{align}
If $N_{kmnq}=\cO(N^0)$, this corresponds in the bulk to a small number
of 1/8-BPS supergravitons in the undeformed \AdSxS\ background
(represented by the $\ket{++}_1$ part).  If $N_{kmnq}=\cO(N)$, on the
other hand, it corresponds to a gas of supergravitons with the
background finitely deformed -- this is the superstratum.\footnote{More
precisely, it is a coherent sum of states of the form
\eqref{mebn31Oct22} with different values of $N_{kmnq}$ that corresponds
to a bulk classical solution \cite{Bena:2017xbt, Giusto:2015dfa}.  }

In \eqref{mebn31Oct22}, we considered descendants built on
$\ket{00}_k$, but we can also consider other anti-chiral primaries
$\ket{\psi}_k$ in \eqref{T4_1-part_ch_pr}.  For the dual bulk state to
be describable by a solution of classical supergravity, the
descendant must be bosonic so that the multiplicity $N_{kmn}$ can
be macroscopic.
If $\ket{\psi}_k$ is bosonic ($h-\hb\in\bbZ$), the states
\eqref{1p_1/8_sugrtn1} and \eqref{1p_1/8_sugrtn3} are bosonic.  The
standard, ``tensor-multiplet'' superstrata \cite{Bena:2015bea,
Bena:2017xbt, Ceplak:2018pws} are macroscopic excitations of such modes.
If $\ket{\psi}_k$ is fermionic ($h-\hb\in\bbZ+\half$), the states
\eqref{1p_1/8_sugrtn2} are bosonic.  The ``vector-multiplet''
superstrata discussed in \cite{Ceplak:2022wri, Ceplak:2022pep} (see also
\cite{Martinec:2022okx}) are macroscopic excitations of such modes.

\subsection{Spectral flow}
\label{ss:spectral_flow}

CFTs with $\cN\ge 2$ supersymmetry has  spectral flow symmetry, which
maps a state with $h,j$ to a state with
\begin{align}
 h'&=h+2\eta j+{c\over 6}\eta^2,\qquad
 j'=j+{c\over 6}\eta.
\label{specflow_h,j}
\end{align}
This is for the left-moving sector. We can do spectral flow
transformations in the right-moving sector, with an independent parameter
$\etab$.

We can interpret spectral flow in two ways. (i)~{\it Passive:} By
spectral flow, we can look at the same state in different
sectors. Namely, it gives different viewpoints on the same state. For
example, we can take a state in the NS sector and flow it  to a state in the R sector.  (ii)~{\it Active:} We
can use spectral flow to transform a state in some sector into a
different state in the same sector. For this to be possible, the
transformed state must be an allowed state in the sector that we are in.

As a concrete example of ``passive'' spectral flow, by taking take
$\eta=\etab=\half$, we can map states in the NS-NS sector to ones in the
R-R sector.  Anti-chiral primaries flow to RR ground states, and the
descendants such as \eqref{mebn31Oct22} go to the RR state
\begin{align}
 \left[\ket{++}_1^{\rm R}\right]^{N_0} \prod_{k,m,n,q} 
 \left[(J_{-1}^+)^m(L_{-1}-J_{-1}^3)^n 
\Bigl(G_{-1}^{+,1}G_{-1}^{+,2}+\tfrac{1}{2h^{\rm NS}}(L_{-1}-J^3_{-1})J_{-1}^+\Bigr)^{\!q\,}
 \ket{00}_k^{\rm R}\right]^{N_{kmnq}}
\label{mbnw12Nov22}
\end{align}
where the superscript ``R'' means Ramond sector states.  It is such
Ramond states that are in direct correspondence with the superstrata in
the bulk.

\subsection{The structure of state construction}
\label{ss:structure_state_construct}

There is a key aspect of the structure of the superstratum states
reviewed above \cite{Mathur:2003hj, Giusto:2013bda, Bena:2015bea}.

On each of the $N$ copies of $\cM$ live the generators
of $SU(1,1|2)^2$, and the ``total'' generator is the sum of such
individual generators. For example,
\begin{align}
 L_{-1}=\sum_{r=1}^N (L_{-1})_r,
\end{align}
where $r$ is the copy number.  When the copies are grouped into 
strands, we can write this also as
\begin{align}
 L_{-1}=\sum_{C} (L_{-1})_C,
\end{align}
where $C$ is the strand number and $(L_{-1})_C$ is the sum of
$(L_{-1})_r$ that belong to the $C$-th strand.

The ground state
\begin{align}
 (\ket{++}_1)^N
\end{align}
is annihilated by $L_{-1}$ because each individual generator $(L_{-1})_r$
kills the NS vacuum $\ket{++}_1$ on the $r$th copy.  This corresponds to
the $SU(1,1|2)^2$ isometry of the dual \AdSxS\ background.  Namely, if
$\ket{\rm vac}$ is the bulk \AdSxS\ vacuum, it is killed by the bulk
isometry generator:
\begin{align}
 L_{-1}\ket{\rm vac}=0.\label{eskt1Nov22}
\end{align}

Now consider including a one-particle anti-chiral primary that is not
$\ket{++}_1$, for example,
\begin{align}
 \ket{00}_k\, (\ket{++}_1)^{N-k}.
\label{jiie1Nov22}
\end{align}
If we act on this state with the total generator $L_{-1}$, all the
generators $(L_{-1})_C$ vanish on
$\ket{++}_1$, except for the $(L_{-1})_C$ acting on the non-trivial
strand $\ket{00}_k$.  Namely,
\begin{align}
 L_{-1} \Bigl[\ket{00}_k\, (\ket{++}_1)^{N-k}\Bigr]
 = (L_{-1}\ket{00}_k) \, (\ket{++}_1)^{N-k}.
\end{align}
Here $L_{-1}\ket{00}_k$ means $\ket{00}_k$ acted upon by $(L_{-1})_C$
that lives on that strand, but we omitted the label $C$.
The bulk dual statement is the following. If one excites one
supergraviton dual to $\ket{00}_k$ in the empty \AdSxS, the bulk state
is $a_{k}^\dagger \ket{\rm vac}$, where $a_{k}^\dagger$ is the creation
operator of the supergraviton. If we act with the bulk isometry
generator on this state, we obtain
\begin{align}
 L_{-1}(a_{k}^\dagger \ket{\rm vac})
= [L_{-1},a_{k}^\dagger]\, \ket{\rm vac},\label{ffdq16Dec22}
\end{align}
because of \eqref{eskt1Nov22}.  So, the total generator $L_{-1}$ acts
only on the supergraviton, changing its wavefunction according to the
particle's $SU(1,1|2)$ representation.  

So, what we are doing in \eqref{1p_1/8_sugrtn}--\eqref{mebn31Oct22} can
be stated as follows.  Consider one supergraviton in the \AdSxS\ vacuum,
represented by an anti-chiral primary $\ket{\psi}_k$. Then we act upon
the state with the total generators to generate descendants like
$(J_0^-)^m(L_{-1})^n\ket{\psi}_k$.  Note that, in the bulk, the
\emph{only} available generators are total generators --- they are
isometries. We simply do not know the bulk dual of individual generators
in CFT\@.  However, when there is only one particle is present, the
total generators act only of the strand.  If we have single-particle
states such as $(J_0^-)^m(L_{-1})^n\ket{\psi}_k$, we can start exciting
many such particles -- even a macroscopic number of them.  They are
superstrata.

\subsection{The orbifold background $\AdSxS/\bbZ_p$}
\label{ss:orb_AdS3xS3/Zp}

The orbifold $\AdSxS/\bbZ_p$ geometry plays an important role in the
current paper.  The construction of the bulk $\AdSxS/\bbZ_p$ geometry
starts with the Lunin-Mathur geometry \cite{Lunin:2001jy, Lunin:2002iz,
Taylor:2005db, Kanitscheider:2007wq} with a circular, $p$-wound profile
function (see Appendix \ref{app:sss:AdS3xS3/Zp}).  The metric of the
Lunin-Mathur geometry is
\begin{align}
 ds_{6E}^2={\Sigma\over\sqrt{Q_1Q_5}}\Bigl[-(dt+{pR_y a^2\over \Sigma}\sin^2\theta\,d\phi)^2
 +(dy-{p R_y a^2\over\Sigma}\cos^2\theta\,d\psi)^2
 \Bigr]+{\sqrt{Q_1 Q_5}\over\Sigma}ds^2(\cB)\label{LM_geom_p-wound}
\end{align}
where $ds^2(\cB),\Sigma$ are defined in
 \eqref{def_base_metric_flat_Sigma}, and the parameters
 $Q_1,Q_5,a,p,R_y$ are related to each other by \eqref{aQR_y/Zp}.  If we
 write this in terms of the new variables
\begin{align}
 \phit
 =\phi-{t\over p R_y}
 ,\qquad
 \psit
 &=\psi-{y\over p R_y},
 \label{jfjx15Dec22}
\end{align}
we obtain the six-dimensional metric 
\begin{align}
 ds^2_{6}
 &=\sqrt{Q_1Q_5}
 \left(-{r^2+a^2\over a^2 (p R_y)^2}dt^2+{r^2\over a^2 (p R_y)^2}dy^2
 +{dr^2\over r^2+a^2}
 +d\theta^2+\sin^2\theta\, d\phit^2+\cos^2\theta\, d\psit^2\right)
 \label{AdS3xS3/Zp_metric}
\end{align}
which is manifestly locally $\AdSxS$ with radius $\cR=(Q_1Q_5)^{1/4}$.
The periodic identifications of the original variables,
\begin{align}
 (\phi,\psi,y)
 \cong
 (\phi+2\pi,\psi,y)
 \cong
 (\phi,\psi+2\pi,y)
 \cong
 (\phi,\psi,y+2\pi R_y),
\end{align}
means the following identifications for the new variables,
\begin{align}
 (\phit,\psit,y)
 \cong
 (\phit+2\pi,\psit,y)
 \cong
 (\phit,\psit+2\pi,y)
 \cong
 \left(\phit,\psit-{2\pi\over p},y+2\pi R_y\right).\label{iorb15Dec22}
\end{align}
So, the metric \eqref{AdS3xS3/Zp_metric} represents a $\bbZ_p$ orbifold
of $\AdSxS$ space.  If $p=1$, the space~\eqref{AdS3xS3/Zp_metric} is the
non-orbifold $\AdSxS$.  The $p>1$ case can be obtained from the
$p=1$ case by the replacement $R_y\to p R_y$, which amounts to dividing
the radius of the $y$ circle by $p$.

The non-orbifold $\AdSxS$ has the isometry group $SO(2,2)\times SO(4)
\cong SL(2)_L\times SU(2)_L \times SL(2)_R\times SU(2)_R$.  If we take
the Killing vectors for this isometry group and set $R_y\to p R_y$, we
find the following ``Killing vectors'' in the orbifold $\AdSxS/\bbZ_p$:
\begin{align}
\label{csp16Dec22}
\begin{split}
 L_0&={i p R_y \over 2}(\partial_t+\partial_y),\\
 L_{\pm {1\over p}}
 &=ie^{\pm {i\over  p R_y}(t+y)}
 \biggl[
 -{p R_y\over 2}\biggl({r\over \sqrt{r^2+a^2}}\partial_t+{\sqrt{r^2+a^2}\over r}\partial_y\biggr)
 \pm {i\over 2}\sqrt{r^2+a^2}\,\partial_r
 \biggr],
\\
 J_0^3&=-{i\over 2}(\partial_{\phit}+\partial_{\psit}),\quad
  J_0^\pm ={i\over 2}e^{\pm i(\phit+\psit)}
 (\mp i\partial_\theta+\cot\theta\, \partial_{\phit}-\tan\theta\, \partial_{\psit}),
\end{split} 
\end{align}
along with right-moving counterparts.  These satisfy the $sl(2)$ and
$su(2)$ algebra relations
\begin{align}
\begin{split}
  [L_0,L_{\pm {1\over p}}]&=\mp L_{\pm {1\over p}},\qquad
 [L_{1\over p},L_{-{1\over p}}]=2L_0,\\
 [J^3_0,J^\pm_0]&=\pm J^{\pm}_0,\qquad
 [J^+_0,J^-_0]=2J^3_0
\end{split}
\end{align}
and leave invariant the local form of the metric.  However, these are
not valid Killing vectors of $\AdSxS/\bbZ_p$, because they are not
single-valued under the last identification in \eqref{iorb15Dec22}.
They are single-valued only in the $p$-covering space where we make the
radius of the $y$-circle $p$~times larger.

When there is a supergraviton in $\AdSxS/\bbZ_p$, if we act with
$L_{-{1\over p}},J_0^+$ on its wavefunction of a particle, the result
will not be single-valued, because of the multi-valuedness mentioned
above.  However, if we act $n$ times with $L_{-1}$ and $m$ times with
$J_0^+$, then $L_{-{1\over p}}^n (J_0^+)^m \propto \exp(-i{m+n\over
p}{y\over R_y})$ and the resulting wavefunction is single-valued if
${m+n\over p}\in\bbZ$. This suggests that, even for the orbifold
$\AdSxS/\bbZ_p$, the procedure around \eqref{ffdq16Dec22} of
generating descendants goes through if we use the Killing vectors of the
covering space, \eqref{csp16Dec22}.  We will discuss the CFT counterpart
of this statement in  section
\ref{sec:CFT_view}.

\subsection{Bulk spectral flow}

The bulk version of spectral flow transformation discussed in section
\ref{ss:spectral_flow} is
\begin{subequations} \label{nitr15Dec22}
 \begin{align}
 \phit &= \phi  - (\eta+\etab) {t\over R_y} -(\eta - \etab){y\over R_y},&
 \psit &= \psi  - (\eta-\etab) {t\over R_y} -(\eta + \etab){y\over R_y},
 \end{align}
 or equivalently,
 \begin{align}
 \phit + \psit &= \phi + \psi - 2\eta \,{t+y\over R_y}
  , &
 \phit - \psit &= \phi - \psi - 2\etab\,{t-y\over R_y}
  .
 \end{align}
\end{subequations}

The Lunin-Mathur geometries are in the RR sector.  The CFT dual of the
geometry \eqref{LM_geom_p-wound} is the RR ground state
\begin{align}
 \bigl(\,\ket{++}_p^{\rm R}\,\bigr)^{N\over p},\qquad
 h_{\rm R}^{}={N\over 4},\qquad
 j_{\rm R}^{}={N\over 2p}.
\end{align}
The transformation \eqref{jfjx15Dec22} to bring the geometry into
\eqref{AdS3xS3/Zp_metric} is spectral flow with $\eta=\etab=-{1\over
2p}$.  So, the solution \eqref{AdS3xS3/Zp_metric} is not in the NS
sector (as we would get if we took $\eta=\etab=-{1\over 2}$) but is in a
different sector, which we dub the NS$'$ sector.  The dual CFT state is
\begin{align}
 \bigl(\,\ket{++}_p^{\rm NS'}\,\bigr)^{N\over p},\qquad
 h_{\rm NS'}^{}={N\over 4}\left(1-{1\over p^2}\right),\qquad
 j_{\rm NS'}^{}=0.
\end{align}

In passing, we mention that, in the R coordinates $(\phi,\psi,y)$, the
generators \eqref{csp16Dec22} have
\begin{align}
 L_{\pm {1\over p}} \propto e^{\pm {i\over  p R_y}(t+y)},\qquad
 J_0^{\pm}\propto e^{\pm i\left(\phi+\psi-{t+y\over pR_y}\right)}
\end{align}
So, the non-single-valuedness of those ``Killing vectors'' is related to
the fractional energy and $y$-momentum in the R sector.

\medskip
The spectral flow transformation \eqref{nitr15Dec22} discussed above is
``passive'' since we take a particular solution and map it into a
solution in a different sector.  The periodicity of the new variables
$(\phit,\psit,y)$ is derived from and different from that of
$(\phi,\psi,y)$.  On the other hand, ``active'' spectral flow
transformations take a solution in a sector and transform it into a
different solution in the same sector.  This is executed by replacing
\begin{align}
 \phi &\to \phi  - (\eta+\bar\eta) {t\over R_y} -(\eta - \bar\eta){y\over R_y},
 &
 \psi &\to \psi  - (\eta-\bar\eta) {t\over R_y} -(\eta + \bar\eta){y\over R_y}
\label{ieyi18Dec22}
\end{align}
in the solution, with the periodicity of $(\phi,\psi)$ unchanged.  If
$\eta,\etab$ are integers, this transformation is an $SL(2,\bbZ)^2$
basis change of coordinate identifications. However, if $\eta,\etab$ are
rational numbers, as we do in section \ref{sec:fract_spectral_flow},
this is an $SL(2,\mathbb{Q})^2$ transformation that changes the physical
solution by changing the orbifold singularity structure.

\section{CFT view on superstrata on $\AdSxS/\bbZ_p$}

\label{sec:CFT_view}

In the construction reviewed in section \ref{s:strata_in_D1-D5_CFT}, it
was crucial that, when we act with total generators on a single
supergraviton state in \AdSxS, it only acts on the supergraviton,
because $\ket{++}_1$ is killed by all rigid generators.
Here we show that this construction can be generalized to the case where
all strand lengths are multiples of a common integer $p$.  One example
is
\begin{align}
 \ket{00}_{kp}\, (\ket{++}_p)^{{N\over p}-k}.
\label{jjmu1Nov22}
\end{align}
In a $p$-covering space,  $\ket{++}_p$ plays the role of
$\ket{++}_1$ because it is annihilated by certain modes defined in the
covering space.
This observation leads to an interesting view on the construction of
superstrata on $\AdSxS/\bbZ_p$.

\subsection{Currents and modes in the covering space}

Let us consider the case where the lengths of all strands are 
integer multiples of $p$, like in~\eqref{jjmu1Nov22}, and focus on one
particular strand with length $kp$.  On this
strand, we have $kp$ copies of currents cyclically glued together, in
the R sector, as
\begin{align}
 T_r(w+2\pi) = T_{r+1}(w),\quad 
 G_r(w+2\pi) = G_{r+1}(w),\quad 
 J_r(w+2\pi) = J_{r+1}(w),
\end{align}
where $r=1,\dots,kp$ is the copy index and $w$ is a cylinder coordinate
with $\Re w\in[0,2\pi)$.  To avoid clutter, we suppress the indices on
$G^{\alpha A},J^{i}$ for now. It is understood that $T_{kp+1}\equiv T_1$,
$G_{kp+1}\equiv G_1$, $J_{kp+1}\equiv J_1$.
If we go to the plane coordinate $z=e^{-iw}$, the gluing condition
 becomes
\begin{align}
 T_r(e^{2\pi i}z) = T_{r+1}(z),\qquad 
 G_r(e^{2\pi i}z) = -G_{r+1}(z),\qquad 
 J_r(e^{2\pi i}z) = J_{r+1}(z),\label{gzsa19Jun22}
\end{align}
where ${\rm arg}\, z\in[0,2\pi)$.

\begin{figure}[tbp]
\begin{quote}
  \begin{center}
  \begin{tikzpicture}
  \scalebox{0.8}{
  \draw[very thick] (0,-0.2) -- +(0,0.4);
  \draw (0,0) -- +(1,0) node [midway,above] {$T_1$};
  \draw (1,-0.2) -- +(0,0.4);
  \draw (1,0) -- +(1,0) node [midway,above] {$T_2$};
  \draw (2,-0.2) -- +(0,0.4);
  \draw (2,0) -- +(2,0) node [midway,above] {$\cdots$};
  \draw (4,-0.2) -- +(0,0.4);
  \draw (4,0) -- +(1,0) node [midway,above] {$T_p$};
  \draw [thick,decorate,decoration={brace,amplitude=10pt,mirror},xshift=0.4pt]
    (0.1,-0.5) -- +(4.8,0) node[midway,below,yshift=-10] {$\Tt_1$};
  \draw[very thick] (5,-0.2) -- +(0,0.4);
  \draw (5,0) -- +(1,0) node [midway,above] {$T_{p+1}$};
  \draw (6,-0.2) -- +(0,0.4);
  \draw (6,0) -- +(1,0) node [midway,above] {$T_{p+2}$};
  \draw (7,-0.2) -- +(0,0.4);
  \draw (7,0) -- +(2,0) node [midway,above] {$\cdots$};
  \draw (9,-0.2) -- +(0,0.4);
  \draw (9,0) -- +(1,0) node [midway,above] {$T_{2p}$};
  \draw[very thick] (10,-0.2) -- +(0,0.4);
  \draw [thick,decorate,decoration={brace,amplitude=10pt,mirror},xshift=0.4pt]
    (5.1,-0.5) -- +(4.8,0) node[midway,below,yshift=-10] {$\Tt_2$};
  \draw (10,0) -- +(3,0) node [midway,above] {$\cdots$};
  \draw (10,0) -- +(3,0) node [midway,below,yshift=-30] {$\cdots$};
  \draw[very thick] (13,-0.2) -- +(0,0.4);
  \draw (13,0) -- +(1,0);
  \draw (14,-0.2) -- +(0,0.4);
  \draw (14,0) -- +(1,0);
  \draw (15,-0.2) -- +(0,0.4);
  \draw (15,0) -- +(2,0)  node [midway,above] {$\cdots$};
  \draw (17,-0.2) -- +(0,0.4);
  \draw (17,0) -- +(1,0) node [midway,above] {$T_{kp}$};
  \draw[very thick] (18,-0.2) -- +(0,0.4);
  \draw [thick,decorate,decoration={brace,amplitude=10pt,mirror},xshift=0.4pt]
    (13.1,-0.5) -- +(4.8,0) node[midway,below,yshift=-10] {$\Tt_k$};
  }
  \end{tikzpicture}
   \vspace*{-1cm}
 \caption{\sl We group $kp$ currents $T_r$ into $k$ groups of $p$ and define
 $T_\rt$.\label{fig:gluing}}
 \end{center}
\end{quote}
\end{figure}
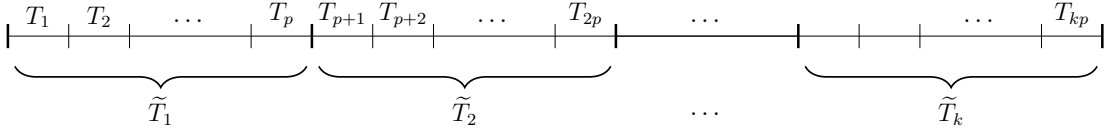

We would like to consider a $p$-covering of this (note that it
is \emph{not} a $kp$-covering). For that, let us group the above $kp$
currents into $k$ groups of $p$, and define $k$ currents $\Tt_{\rt}(w)$,
$\rt=1,\dots,k$, as follows:
\begin{align}
\begin{aligned}
  \Tt_1(w)&=T_1(w),    & \Tt_1(w+2\pi)&=T_2(w),    &\cdots,&& \Tt_1(w+2(p-1)\pi)&=T_p(w),\\
  \Tt_2(w)&=T_{p+1}(w),& \Tt_2(w+2\pi)&=T_{p+2}(w),&\cdots,&& \Tt_2(w+2(p-1)\pi)&=T_{2p}(w),\\
 &\cdots,\\
  \Tt_k(w)&=T_{(k-1)p+1 }(w),& \Tt_k(w+2\pi)&=T_{(k-1)p+2}(w),&\cdots,&& \Tt_k(w+2(p-1)\pi)&=T_{kp}(w).\\
\end{aligned}
\end{align}
We likewise define $\Gt_\rt(w),\Jt_\rt(w)$.
These currents live on a $p$-cover of the cylinder/plane with $\Re w\in[0,2\pi p)$ or ${\rm arg}\, z\in[0,2\pi p)$.
See Figure \ref{fig:gluing}.
The periodicity of these currents is, on the cylinder,
\begin{align}
 \Tt_\rt(w+2\pi p)&= \Tt_{\rt+1}(w),&
 \Gt_\rt(w+2\pi p)&= \Gt_{\rt+1}(w),&
 \Jt_\rt(w+2\pi p)&= \Jt_{\rt+1}(w),
\end{align}
and, on the plane,
\begin{align}
 \Tt_\rt(e^{2\pi  p i} z)&= \Tt_{\rt+1}(z),&
 \Gt_\rt(e^{2\pi  p i} z)&= (-1)^p\Gt_{\rt+1}(z),&
 \Jt_\rt(e^{2\pi  p i} z)&= \Jt_{\rt+1}(z).
\end{align}
If we define the plane coordinate $t$ on a $p$-covering space by
$t=z^{1/p}$, ${\rm arg}\,t\in[0,2\pi)$, these currents have the
following periodicity:
\begin{align}
 \Tt_\rt(e^{2\pi i} t)&= \Tt_{\rt+1}(t),&
 \Gt_\rt(e^{2\pi i} t)&= -\Gt_{\rt+1}(t),&
 \Jt_\rt(e^{2\pi i} t)&= \Jt_{\rt+1}(t).
\end{align}
Comparing this with \eqref{gzsa19Jun22}, we see that we are in the R sector on the $t$ plane.

Let us define the total currents
\begin{align}
 T(z)\equiv \sum_{r=1}^{kp} T_r(z),\qquad
 G(z)\equiv \sum_{r=1}^{kp} G_r(z),\qquad
 J(z)\equiv \sum_{r=1}^{kp} J_r(z).
\end{align}
These are the standard currents of the CFT that define conserved charges
such as $h,j$.   They have the periodicity
\begin{align}
 T(e^{2\pi i }z)=T(z),\qquad
 G(e^{2\pi i }z)=-G(z),\qquad
 J(e^{2\pi i }z)=J(z),\label{dnlo19Jun22}
\end{align}
as is clear from the periodicity of the individual currents,
\eqref{gzsa19Jun22}.  Also, they satisfy the standard OPE relations
\begin{align}
 T(z_1) T(z_2) \sim {c_{k p}\over 2z_{12}^4}+{2T(z_2)\over z_{12}^2}+{\p T(z_2)\over z_{12}},\qquad etc.
\end{align}
with $z_{12}=z_1-z_2$ and $c_n\equiv n c_1$, where $c_1$ is the central
charge for a single copy. For the D1-D5 CFT, $c_1=6$. These OPEs can be
easily proven by the fact that the individual currents $T_r(z)$ etc.\
satisfy the OPEs with central charge $c_1$.

We also define
\begin{align}
 \Tt(z)\equiv \sum_{\rt=1}^{k} \Tt_\rt(z),\qquad
 \Gt(z)\equiv \sum_{\rt=1}^{k} \Gt_\rt(z),\qquad
 \Jt(z)\equiv \sum_{\rt=1}^{k} \Jt_\rt(z).
\end{align}
These are something we need because we want to consider a $p$-cover,
even though our strand has length $kp$.  These can be regarded as the
total currents on the $p$-cover. Their periodicity is
\begin{align}
 \Tt(e^{2\pi i p}z)=\Tt(z),\qquad
 \Gt(e^{2\pi i p}z)=(-1)^p\Gt(z),\qquad
 \Jt(e^{2\pi i p}z)=\Jt(z)\label{dnqe19Jun22}
\end{align}
and their OPE is
\begin{align}
 \Tt(z_1) \Tt(z_2) \
 \sim {c_{k}\over 2z_{12}^4}+{2\Tt(z_2)\over z_{12}^2}+{\p \Tt(z_2)\over z_{12}},\qquad etc,\qquad
 {\rm arg}\,z_{1,2}\in[0,2\pi p).\label{hcme19Jun22}
\end{align}
Note also that
\begin{align}
 T(z)=\sum_{l=0}^{p-1}\Tt(e^{2\pi i l}z),\qquad 
 G(z)=\sum_{l=0}^{p-1}\Gt(e^{2\pi i l}z),\qquad 
 J(z)=\sum_{l=0}^{p-1}\Jt(e^{2\pi i l}z).
\label{doen19Jun22}
\end{align}

The periodicity \eqref{dnlo19Jun22} of the
total currents $T(z),G(z),J(z)$
means that their $z$ expansions are integer-moded:
\begin{align}
 T(z)=\sum_{n\in\bbZ} {L_n^{(z)}\over z^{n+2}},\qquad
 G(z)=\sum_{n\in\bbZ} {G_n^{(z)}\over z^{n+{3\over 2}}},\qquad
 J(z)=\sum_{n\in\bbZ} {J_n^{(z)}\over z^{n+1}}.\label{hlnn19Jun22}
\end{align}
We put the superscript ``$(z)$'' on $L_n^{(z)}$ etc.~to clarify that
they are the expansion coefficients in the $z$ coordinate.  On the other
hand, \eqref{dnqe19Jun22} means that the $z$ expansions of the
covering-space total currents $\Tt(z),\Gt(z),\Jt(z)$ are $1\over
p$-moded as
\begin{align}
 \Tt(z)=\sum_{n\in\bbZ} {\Lt_{n\over p}^{(z)}\over z^{{n\over p}+2}},\qquad
 \Gt(z)=\sum_{n\in\bbZ} {\Gt_{n\over p}^{(z)}\over z^{{n\over p}+{3\over2}}},\qquad
 \Jt(z)=\sum_{n\in\bbZ} {\Jt_{n\over p}^{(z)}\over z^{{n\over p}+1}}.
\end{align}
The relation \eqref{doen19Jun22} means that
\begin{align}
 L_n^{(z)}=p\Lt^{(z)}_n,\qquad
 G_n^{(z)}=p\Gt^{(z)}_n,\qquad
 J_n^{(z)}=p\Jt^{(z)}_n,\qquad n\in\bbZ.\label{hmig19Jun22}
\end{align}
Let us also expand the $t$-space currents $\Tt(t),\Gt(t),\Jt(t)$ as
\begin{align}
 \Tt(t)=\sum_{n\in\bbZ} {\Lt_{n}^{(t)}\over t^{n+2}},\qquad
 \Gt(t)=\sum_{n\in\bbZ} {\Gt_{n}^{(t)}\over t^{n+{3\over2}}},\qquad
 \Jt(t)=\sum_{n\in\bbZ} {\Jt_{n}^{(t)}\over t^{n+1}}.
\end{align}
Because the $\Tt\Tt$ OPE has central charge $c_{k}$ as in
\eqref{hcme19Jun22}, the coordinate transformation between
$(\Tt(z),\Gt(z),\Jt(z))$ and $(\Tt(t),\Gt(t),\Jt(t))$ is
\begin{align}
 (t')^2\Tt(t)&=\Tt(z)-{c_k\over 12}\{t,z\},\qquad
 (t')^{3/2}\Gt(t)=\Gt(z),\qquad
 t'\Jt(t)=\Jt(z),
\end{align}
where $t'=\partial_z t$ and $\{\,,\,\}$ is the Schwarzian. From this, we
find that the relation between the modes of $\Tt,\Gt,\Jt$ in the $z$
coordinate and those in the $t$ coordinate is
\begin{align}
 \Lt^{(t)}_n=p^2\Lt^{(z)}_{n\over p}-{c_k(p^2-1)\over 24}\delta_{n0},\qquad
 \Gt^{(t)}_n=p^{3\over 2}\Gt^{(z)}_{n\over p},\qquad
 \Jt^{(t)}_n=p\Jt^{(z)}_{n\over p}.\label{hmop19Jun22}
\end{align}

\subsection{Lifting anti-chiral primaries to the covering space}

The construction of supergraviton states was based on the following: if
we have an anti-chiral primary with $h=-j$ in the NS sector, we can
generate the spectrum of the $SU(1,1|2)$ multiplet by repeatedly acting
on it with the modes of $T,G,J$, namely $J^+_0,G^{+,A}_{-1/2},L_{-1}$.
Here we show that, for an anti-chiral primary on a strand of length $kp$,
there is a different way to generate a different ``spectrum''.  Namely,
if we go to an ``NS sector'' in the covering $t$ space, the state is an anti-chiral primary with respect to $\Tt,\Gt,\Jt$ and we can generate
descendants by the action of their coefficients.

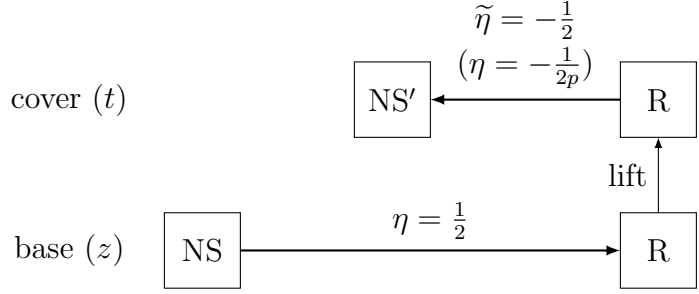
\begin{figure}
 \begin{quote}
 \begin{center}
  %
 \begin{tikzpicture}
 \draw (-6,0) rectangle node{NS} +(-1,1);
 \draw[thick,latex-] (-1,0.5) -- +(-5,0) node [above,midway] {$\eta=\half$};
 \draw (0,0) rectangle node{R} +(-1,1);
 \node () at (-8.25,0.5) {base $(z)$};
 \draw[-latex] (-0.5,1) -- +(0,1) node [left,midway] {lift};
 \draw (0,2) rectangle node{R} +(-1,1);
 \node () at (-8.25,2.5) {cover $(t)$};
 \draw[thick,-latex] (-1,2.5) -- +(-2.5,0) node [above,midway] 
   {$\begin{array}{c}\etat=-\half \\[.5ex] (\eta=-{1\over 2p})\end{array}$};
 \draw (-3.5,2) rectangle node{NS$'$} +(-1,1);
 \end{tikzpicture}
  \caption{\label{fig:flows} \sl Fractional spectral flows in the base and
  covering spaces}
 \end{center}
 \end{quote}
\end{figure}

Let us start with a state with $(L^{(z)}_0,J^{(z)}_0)=(h,j)$ in the
NS sector.  Note that we are using the modes of the total currents,
\eqref{hlnn19Jun22}, which define the standard charges of CFT\@.  To be
very clear (although it's a lot of scripts), we can display the sector
and write
\begin{align}
 (L^{(z),\rm NS}_0,J^{(z),\rm NS}_0)=(h,j).
\end{align}
Now we
go to the R sector by $\eta=\half,c=c_{kp}=kpc_1$.  We find that
\begin{align}
 (L^{(z),\rm R}_0,J^{(z),\rm R}_0)
=\left(h+j+{pkc_1\over 24},\,j+{pkc_1\over 12}\right).
\end{align}
If we translate this to the modes of the covering-space total currents
$\Tt(z),\Jt(z)$, using \eqref{hmig19Jun22}, we find
\begin{align}
 (\Lt^{(z),\rm R}_0,\Jt^{(z),\rm R}_0)
 =\left({h+j\over p}+{kc_1\over 24},\,{j\over p}+{kc_1\over 12}\right).
\end{align}
If we go to the $t$ coordinate using \eqref{hmop19Jun22}, we find
\begin{align}
 (\Lt^{(t),\rm R}_0,\Jt^{(t),\rm R}_0)
 =\left(p(h+j)+{kc_1\over 24},\,j+{kpc_1\over 12}\right).
\end{align}
Finally, we spectral flow in the $t$ coordinate, by $\etat=-\half$,
$c=c_k=kc_1$ (we use $\etat$ for the parameter in the $t$ space; this
corresponds to $\eta={\etat\over p}=-{1\over 2p}$ in the base $z$ space).
Let us call the resulting sector the ``NS$'$ sector''.  Thus we find the
following charges in the $p$-cover:
\begin{align}
 (\Lt^{(t),\rm NS'}_0,\Jt^{(t),\rm NS'}_0)
 =\left(p(h+j)-j+(1-p){kc_1\over 12},\, j+(p-1){kc_1\over 12}\right)
\equiv (h',j').
\label{neuq2Nov22}
\end{align}
See Figure \ref{fig:flows} for a schematic diagram of the fractional
spectral flows in the base and covering spaces.

The relation \eqref{neuq2Nov22} means that, if we start with an anti-chiral
primary with $h=-j$, the charges of the final state are
\begin{align}
 h'=-j'=h+(1-p){kc_1\over 12}.
\end{align}
Namely, it is an anti-chiral primary  again, but now with respect
to the algebra generated by the covering-space currents
$\Tt(t),\Gt(t),\Jt(t)$.
In the present case of the D1-D5 CFT, anti-chiral primaries
\eqref{T4_1-part_ch_pr} with length $kp$ in the base space lift to
anti-chiral primaries with length $k$ in the covering space as follows:
\begin{align}
\renewcommand{\arraystretch}{1.8}
\begin{array}{c|c||c|cc}
 \hline
 \begin{minipage}{12ex}anti-chiral primary\end{minipage}
 &
 L^{(z),\rm NS}_0=J^{(z),\rm NS}_0
 &
 \Lt^{(t),\rm NS'}_0=\Jt^{(t),\rm NS'}_0
 &
 \begin{minipage}{22ex}anti-chiral primary in the covering space\end{minipage}
 \\
 \hline\hline
 \ket{\alpha}_{kp}&h=j={kp-\alpha\over 2}&h'=j'={k-\alpha \over 2} & \ket{\alpha }_k\\
 \hline
 \ket{\dot{A}}_{kp}  &h=j={kp \over 2}    &h'=j'={k\over 2}      & \ket{\dot{A}}_k\\
 \hline
\end{array}
\label{juya19Jun22}
\end{align}
Here we only considered the left-moving part of the anti-chiral
primaries \eqref{T4_1-part_ch_pr}; we can get full anti-chiral primaries
by multiplying left and right-moving parts.  We see that an anti-chiral
primary in the base simply lifts to the \emph{same} anti-chiral primary,
only with its length shortened by a factor of $p$.  In particular, the
strand in the base
\begin{align}
 \ket{++}_p
\end{align}
lifts in the covering space to the NS$'$ vacuum $\ket{++}_1$ with
$h'=j'=0$.  This is annihilated by the rigid generators of
$\Tt(t),\Gt(t),\Jt(t)$.

\subsection{The new spectrum of descendants}

Because we have (NS$'$) anti-chiral primaries, just as we did in the
base space, we can start generating descendants by acting on them by the
modes of $\Tt(t),\Gt(t),\Jt(t)$, namely
\begin{align}
 \Lt_{-1}^{(t),{\rm NS'}},\quad
 \Gt_{-\half}^{(t),{\rm NS'},+A},\quad
 \Jt_0^{(t),{\rm NS'},+}.\label{june19Jun22}
\end{align}
The number of times we can act with these generators is the same as in
the non-orbifold case, if we replace the base-space quantum numbers
$h=-j$ with the covering space quantum numbers $h'=-j'$.  

More explicitly, if we start in the base space with a single-particle NS
anti-chiral primary~$\ket{\psi}_{pk}^{\rm NS}$, it lifts, via the
process of Figure \ref{fig:flows}, to the single-particle NS$'$
anti-chiral primary~$\ket{\psi}_k^{\rm NS'}$.  Building on this, in
exactly the same way as \eqref{1p_1/8_sugrtn}, we can generate
one-particle descendant states:
\begin{subequations} 
 \begin{align}
 \bigl(\Jt_0^{(t),{\rm NS'},+}\bigr)^m\,
 \bigl(\Lt_{-1}^{(t),{\rm NS'}}\bigr)^n\,
 &\ket{\psi}_k^{\rm NS'},\\
 \bigl(\Jt_0^{(t),{\rm NS'},+}\bigr)^m\,
 \bigl(\Lt_{-1}^{(t),{\rm NS'}}\bigr)^n\,
 \Gt_{-\half}^{(t),{\rm NS'},+,A}
 &\ket{\psi}_k^{\rm NS'},
 \\
 \bigl(\Jt_0^{(t),{\rm NS'},+}\bigr)^m\,
 \bigl(\Lt_{-1}^{(t),{\rm NS'}}\bigr)^n\,
 \left(\Gt_{-\half}^{(t),{\rm NS'},+,1}\Gt_{-\half}^{(t),{\rm NS'},+,2}
 +\tfrac{1}{2h'}\Lt_{-1}^{(t),{\rm NS}}\Jt_0^{(t),{\rm NS'},-}\right)
  &\ket{\psi}_k^{\rm NS'}.
 \end{align}
\end{subequations}

We can repeat the procedure of constructing states in \AdSxS\ discussed
in section \ref{ss:structure_state_construct}.  For example, let us
start in the base space with
\begin{align}
\ket{00}_{pk}^{\rm NS}
 \bigl(\ket{++}_{p}^{\rm NS}\bigr)^{{N\over p}-k},
\end{align}
which represents a single-particle supergraviton in empty
$\AdSxS/\bbZ_p$.  This lifts to the state in the covering space,
\begin{align}
\ket{00}_{k}^{\rm NS'}
 \bigl(\ket{++}_1^{\rm NS'}\bigr)^{{N\over p}-k}.\label{vfb12Nov22}
\end{align}
If we act on this $n$ times with the total generator $\Lt_{-1}^{(t),{\rm
NS'}}$, we get
\begin{align}
(\Lt_{-1}^{(t),{\rm NS'}})^n\!
 \left[\ket{00}_{k}^{\rm NS'}
 \Bigl(\ket{++}_1^{\rm NS'}\Bigr)^{{N\over p}-k}\right]
 =
 \Bigl((\Lt_{-1}^{(t),{\rm NS'}})^n\ket{00}_{k}^{\rm NS'}\Bigr)\,
 \Bigl(\ket{++}_1^{\rm NS'}\Bigr)^{{N\over p}-k}.\label{fgvx12Nov22}
\end{align}
On the right, $\Lt_{-1}^{(t),{\rm NS'}}$ is the individual generator
living on the strand $\ket{00}_{k}^{\rm NS'}\!$. This is the only
surviving operator, because all other individual generators in the total
generator $\Lt_{-1}^{(t),{\rm NS'}}$ annihilates the ``vacuum strand''
$\ket{++}_1^{\rm NS'}$.  In the bulk, we are acting with a ``Killing
vector''~\eqref{csp16Dec22} of $\AdSxS/\bbZ_p$, which transforms the
wavefunction of the single-particle supergraviton state.  Recall that
the ``Killing vectors'' of $\AdSxS/\bbZ_p$ are really the Killing
vectors of the $p$-covering space, namely $\AdSxS$. So, the structure of
state generation is completely parallel in CFT and the bulk.

If we translate this into the R sector in the base space, the right-hand
side of \eqref{fgvx12Nov22} becomes
\begin{align}
 \Bigl(\bigl(\Lt_{-{1\over p}}^{(z),{\rm R}}\bigr)^n\ket{00}_{pk}^{\rm R}\Bigr)\,
 \Bigl(\ket{++}_{p}^{\rm R}\Bigr)^{{N\over p}-k}.
\end{align}
This is an allowed state if ${n\over p}\in\bbZ$.  In the bulk, this
requires that the wavefunction of the supergraviton be single-valued.
If we excite multiple such supergravitons, we will get a state like
\begin{align}
 &\left[\ket{++}_{p}^{\rm R}\right]^{N_0} 
 \prod_{k,m,n,q}
 \biggl[
 \Bigl(\Jt_{-{1\over p}}^+\Bigr)^m
 \Bigl(\Lt_{-{1\over p}}-\frac{1}{p}\Jt_{-{1\over p}}^3\Bigr)^n 
 \notag\\[-1ex]
 &\hspace*{24ex}
 \biggl(\Gt_{-{1\over p}}^{+,1}\Gt_{-{1\over p}}^{+,2}+\frac{1}{2h'^{\rm NS}}\Bigl(\Lt_{-{1\over p}}-\frac{1}{p}\Jt^3_{-{1\over p}}\Bigr)\Jt_{-{1\over p}}^+\biggr)^{\!q\,}
\ket{00}_{kp}^{\rm R}\biggr]^{N_{kmnq}}  
\end{align}
where $pN_0+\sum_{k,m,n,q}k p N_{kmnq}=N$.  This is a superstratum in
$\AdSxS/\bbZ_p$, generalizing the ordinary superstrata
\eqref{mbnw12Nov22} in the non-orbifold \AdSxS\ background.  For this
state to be a valid state of the symmetric orbifold CFT, we must only include
$(k,m,n,q)$ with
\begin{align}
 {m+n+2q\over p}\in\bbZ.\label{ihab16Dec22}
\end{align}

\section{Gravity side}
\label{sec:gravity_side}

In the previous section, we explained in detail the CFT understanding of
the construction of superstrata on $\AdSxS/\bbZ_p$.  In this section, we
will turn to the gravity aspects of superstrata on $\AdSxS/\bbZ_p$.  We
will write down some simple examples of superstrata on $\AdSxS/\bbZ_p$
and analyze their geometries.  Because the bulk superstrata are
naturally in the R sector, the corresponding CFT states are also in the
R sector.

\subsection{The $(k,0,0)_{p}$ superstratum}
\label{ss:(k,0,0)_p}

First, although this is not really a superstratum, consider the 1/4-BPS
state
\begin{align}
 (\ket{++}_p)^{N_0} (\ket{00}_{kp})^{N^{00}_{kp}} ,\qquad
 N_0 + kN^{00}_{kp}={N\over p}.\label{ftwj19Dec22}
\end{align}
We can call this a $(k,0,0)_p$ superstratum.  The gravity solution dual
to this state is found in Appendix \ref{app:sss:(k,0,0)_p}, and it is a
deformation of the $\AdSxS/\bbZ_p$ solution, \eqref{AdS3xS3/Zp_metric},
by a parameter~$b$.

We can write the 6D metric as
\begin{align}
 ds^2_{6E}&=G_{tt}dt^2+G_{yy}dy^2+G_{rr}dr^2+G_{\theta\theta} d\theta^2\notag\\
 &\quad +G_{\phi\phi}(d\phi+A^\phi_t dt)^2
 +G_{\psi\psi}(d\psi+A^\psi_y dy)^2,
\end{align}
where
\begin{align}
 G_{tt}&=-{(r^2+a^2)\Sigma \sqrt{f}\over pR_y((r^2+a^2)f-a^4\sin^2\theta)},\quad
 G_{yy}={r^2\Sigma \sqrt{f}\over pR_y(r^2 f+a^4\cos^2\theta)},\quad
 G_{rr}={pR_y\sqrt{f}\over r^2+a^2},\notag\\
 A_\phi&=-{a^2 \Sigma \over pR_y((r^2+a^2)f-a^4\sin^2\theta)},\quad
 A_\psi=-{a^2 \Sigma \over pR_y(r^2f+a^4\cos^2\theta)},\\
f&\equiv a^2+{b^2\over 2}-{b^2a^{2k}\sin^{2k}\theta\over 2(r^2+a^2)^k}.\notag
\end{align}
Let us assume that $a\sim b$. At large distance $r\gg a\sim b$, we have
\begin{align}
 f\approx a^2+{b^2\over 2}={\cR^4\over (pR_y)^2},
\qquad\cR\equiv (Q_1 Q_5)^{1/4},
\end{align}
and
\begin{align}
 G_{yy}\approx {r^2\over \cR^2},\qquad G_{tt}\approx -{r^2\over \cR^2},\qquad
 G_{rr}\approx {\cR^2\over r^2}.
\end{align}
So, the $(t,r,y)$ part is AdS$_3$ with radius $\cR=(Q_1Q_5)^{1/4}$.

The undeformed $\AdSxS/\bbZ_p$ solution had a $\bbZ_p$ singularity at
$r=0,\theta=\pi/2$.  In the present solution, near that point, we find
that
\begin{align}
\begin{gathered}
 G_{yy}\approx {r^2\over (pR_y)a},\quad
 G_{tt}\approx -{r^2\over (pR_y)a},\quad
 G_{rr}\approx {pR_y\over a},\quad
 A_\phi \approx -{dt\over pR_y},\quad
 A_\psi \approx -{dy\over pR_y},\quad
\end{gathered}
\end{align}
and there is no effect of $b$.  Therefore the structure of the orbifold
singularity is exactly the same as the $b=0$ case, $\AdSxS/\bbZ_p$.  We
can see this by the spectral flow \eqref{jfjx15Dec22} we used for the
empty $\AdSxS/\bbZ_p$ background.

\bigskip
More generally, we can see that superstratum fluctuations do not change
the structure of the $\bbZ_p$ orbifold. 

By setting $R_y \to p R_y$ in the superstratum fluctuation around the
non-orbifold $\AdSxS$~\cite{Bena:2017xbt}, we get the fluctuation around
the orbifold $\AdSxS/\bbZ_p$. We find
\begin{align}
\label{jhmr18Dec22}
\begin{split}
  Z_4&= pR_y
 {\Delta_{kmn}\over \Sigma} e^{iv_{kmn}},\\
 \Theta_4&=
 -\sqrt{2}\,
 \Delta_{kmn}
 \biggl[i\left(-(m+n)r\sin\theta +n\left(1-{m\over k}\right){\Sigma\over r \sin\theta}  \right)\Omega^{(1)}\\
 &\qquad \qquad \qquad \qquad 
 +m\left(1+{n\over k}\right)\Omega^{(2)} +\left({m\over k}-1\right)n\, \Omega^{(3)} \biggr]
 e^{iv_{kmn}'}
\end{split}
\end{align}
where $k,m,n\ge 0$ are integers, and $v_{kmn}'\equiv v_{kmn}|_{R_y\to p
R_y}$.  The quantities $\Delta_{kmn},v_{kmn}$ are defined
in~\eqref{Delta_v_kmn_def}.  We have the phase factor
\begin{align}
e^{iv_{kmn}'}
 =\exp\Bigl[i\Bigl((m+n){t+y\over pR_y}+(k-m)\phi -m\psi\Bigr)\Bigr].
\label{jicf18Dec22}
\end{align}
For this to be single-valued as $y\to y+2 \pi R_y$, we need ${m+n\over
p}\in\bbZ$.  This is the bulk manifestation of the requirement
\eqref{ihab16Dec22}.

To study the behavior of fields near the singularity $r=0,\theta=\pi/2$,
for example, let us consider the scalar~$C_0$.  It goes like
\begin{align}
 C_0={Z_4\over Z_1}\propto
\Delta_{kmn}\, e^{iv_{kmn}'}
 \sim
 r^n 
 {\sin^{k-m}\!\theta\,\cos^m\!\theta}\,e^{iv_{kmn}}.
\end{align}
If $n>0$ or $m>0$, this vanishes at the singularity and is harmless.  If
$n=m=0$, it is finite on the singularity,
\begin{align}
 C_0={Z_4\over Z_1}
 \sim
 e^{ik\phi}.
\end{align}
Because the $\phi$ circle remains finite at the singularity, this $\phi$
dependence does not lead to any additional singularity.

So, we expect that superstrata do not change the $\bbZ_p$ singularity
structure of the  orbifold $\AdSxS/\bbZ_p$ background.

\subsection{The $(1,0,n)_p$ stratum}

The original, non-orbifold $(k,m,n)$ superstratum represents the CFT
state
\begin{align}
 \Bigl[\ket{++}_1\Bigr]^A
 \Bigl[(J_{-1}^+)^m (L_{-1}-J^3_{-1})^n\ket{00}_k\Bigr]^B.\label{mbnf12Nov22}
\end{align}
We want to consider the orbifold version of this, namely a $(k,m,n)_p$
superstratum, whose CFT dual is
\begin{align}
 \Bigl[\ket{++}_p\Bigr]^A
 \Bigl[(\Jt_{-{1\over p}}^+)^m\left(\Lt_{-{1\over p}}-\tfrac{1}{p}\Jt^{3}_{-{1\over p}}\right)^n\ket{00}_{kp}\Bigr]^B,
\qquad {m+n\over p}\in \bbZ.\label{iaqw5Jul22}
\end{align}  
The corresponding geometry is obtained by taking the ansatz data for the
$(k,m,n)$ stratum \cite{Bena:2017xbt} and setting $R_y\to p
R_y$. Explicitly, we have
\begin{equation}
 \begin{aligned}
 Z_1&={Q_1\over \Sigma}+{b_4^2\,(p R_y)^2\over 2Q_5}\,{\Delta_{2k,2m,2n}\over \Sigma}\cos v'_{2k,2m,2n},\\
 Z_2&={Q_5\over \Sigma},\qquad Z_4=b_4 (p R_y) {\Delta_{k,m,n}\over \Sigma}\cos v_{k,m,n}',\\\
 \Theta_1&=0,\qquad \Theta_2={b_4^2(p R_y)\over 2Q_5}\vartheta'_{2k,2m,2n},\qquad
 \Theta_4=b_4 \vartheta'_{k,m,n}
\end{aligned}\label{iasz5Jul22}
\end{equation}
where $v_{m,n,k}'=v_{m,n,k}|_{R_y\to p R_y}$, $\vartheta'_{k,m,n}=
 \vartheta_{k,m,n}|_{R_y\to p R_y}$.
Also,
\begin{equation}
 \begin{aligned}
 {Q_1 Q_5\over (p R_y)^2}&=a^2+{b^2\over 2},\qquad
  b^2=x_{k,m,n}b_4^2,\qquad x_{k,m,n}^{-1}={k\choose m}{k+n-1\choose n}
\end{aligned}\label{pai5Jul22}
\end{equation}
\begin{equation}
 \beta=p  \beta_0,\qquad 
\cF=\cF_{k,m,n},\qquad 
\omega=p(\omega_0+\omega_{k,m,n}).
\end{equation}

Let us see the solution more explicitly, for a $(1,0,n)$ stratum.  In
this case, we have
\begin{equation}
 \begin{gathered}
b=b_4,\qquad \cF_{1,0,n}=-{b_4^2\over a^2}\left(1-{r^{2n}\over (r^2+a^2)^{n}}\right),
  \\
 \omega_{1,0,n}=
 {b_4^2(p R_y)\over\sqrt{2}\,\Sigma}\left(1-{r^{2n}\over (r^2+a^2)^{n}}\right)\sin^2\theta\,d\phi.
\end{gathered}
\end{equation}
Just as in the non-orbifold case \cite{Bena:2017xbt}, we can write the
6D metric as
\begin{align}
\begin{split}
 ds_{6E}^2 =& - \frac{\Lambda}{\sqrt{Q_1 Q_5}} \;\! \frac{2a^2(r^2+a^2)}{2a^2+ b^2 F_0(r)} \;\! dt^2
 \\
 &+ \frac{1}{\sqrt{Q_1 Q_5} \;\! \Lambda} \,
 \frac{ r^2 \left( 2a^2+b^2 F_0(r) \right)  F_2(r,\theta)  }
 {2a^2 \bigl[r^2 (2a^2+b^2) +a^2 (2a^2+b^2 F_0(r))  \bigr]}
 \left( dy + \frac{b^2F_0(r)}{2a^2+b^2F_0(r)} dt \right)^2
 \\
 &+ \sqrt{Q_1 Q_5} \;\! \Lambda \left( \frac{dr^2}{r^2+a^2} + d\theta^2 \right) 
 + \frac{\sqrt{Q_1 Q_5}}{\Lambda} \sin^2\theta \left( d\phi - 
 \frac{2a^2}{2a^2+b^2} \frac{dt}{p R_y} \right)^2 
 \\
 & {} + \frac{\sqrt{Q_1 Q_5}}{\Lambda} { 2a^2+b^2 F_1(r)\over 2a^2+b^2}  \cos^2\theta
 \left( d\psi - \frac{\left(2a^2+b^2F_0(r)\right)dy + b^2 F_0(r) \;\! dt}{\left(2a^2+b^2F_1(r)\right)p R_y} \right)^2
\end{split}
\label{eq:10n-metric}
\end{align}
where  we have used the
shorthand notation
\begin{equation}
\begin{aligned}
 \Lambda 
 = \sqrt{ 1 - \frac{a^2\,b^2}{(2 a^2 +b^2)} \, \frac{r^{2n}}{(r^2 +a^2)^{n+1}} \, \sin^2 \theta  } 
 = \sqrt{ 1 - \frac{b^2}{2 a^2 +b^2} \Delta_{1,0,n}^2  } \,,
  \label{Lambdadef1}
\end{aligned}
\end{equation}
\begin{equation}
\begin{aligned}
  F_0(r) &= 1-\frac{r^{2n}}{(r^2+a^2)^n} \,, \qquad F_1(r) ~=~ 1 - \frac{a^2}{r^2+a^2}\frac{r^{2n}}{(r^2+a^2)^n} \,, \cr
 F_2(r,\theta) &= r^2 (2a^2+b^2) +a^2 \left[ 2a^2 +b^2 \left( 1 - \frac{r^{2n}}{(r^2+a^2)^n}\sin^2\theta \right) \right] .
\end{aligned}
\end{equation}
Some dependence of the metric on $p$ through the replacement $R_y\to p
R_y$ has been absorbed into $\cR^2=\sqrt{Q_1Q_5}=pR_y\sqrt{a^2+b^2/2}$ by \eqref{pai5Jul22} and, as a
result, $p$ enters only in how the coordinates $t,y$ mix with
$\phi,\psi$.  

Let us define 
the parameter 
\begin{align}
 \gamma\equiv {2a^2\over b^2}
\end{align}
that measures the length of the throat region (smaller $\gamma$ means a
longer throat).  In terms of~$\gamma$, the relation \eqref{pai5Jul22}
reads
\begin{align}
 {\cR^4\over (p R_y)^2}&={b^2\over 2}(1+\gamma).
\end{align}
Let us focus on the long throat limit,
\begin{align}
 \gamma \ll 1,\qquad \text{i.e.,}\quad a\ll b.
\end{align}
We can divide the geometry into three
regions,
$a\ll b\ll r$ (region I),
$a\ll r\ll b$ (region II),
and
$r\ll a\ll b$ (region III).  Note that
\begin{align}
 a\ll r: &\qquad \Lambda\approx 1,~ F_0\approx {a^2n\over r^2},~F_2\approx b^2r^2,\qquad
G_{yy}\approx {r^2+nb^2/2\over\sqrt{Q_1 Q_5}},\\
 r\ll a\ll b: &\qquad \Lambda\approx 1,~F_0\approx 1, ~ F_2\approx a^2b^2,\qquad G_{yy}\approx {b^2r^2\over 2\sqrt{Q_1Q_5}a^2}.
\end{align}
So, 
\begin{align}
 G_{yy}\approx 
\begin{cases}
\displaystyle{r^2\over\sqrt{Q_1 Q_5}}={r^2\over\cR^2} & (b\ll r,~\text{region I})\\[3ex]
\displaystyle{nb^2\over 2\sqrt{Q_1 Q_5}}\approx {n\cR^2\over (pR_y)^2} & (a\ll r \ll b,~\text{region II})\\[3ex]
\displaystyle{b^2r^2\over 2\sqrt{Q_1Q_5}a^2}={r^2\over \gamma\cR^2}& (r\ll a,~\text{region III})\\
\end{cases}
\end{align}%
\begin{figure}[tbp]
  \begin{quote}
 \begin{center}
  \includegraphics[height=8cm]{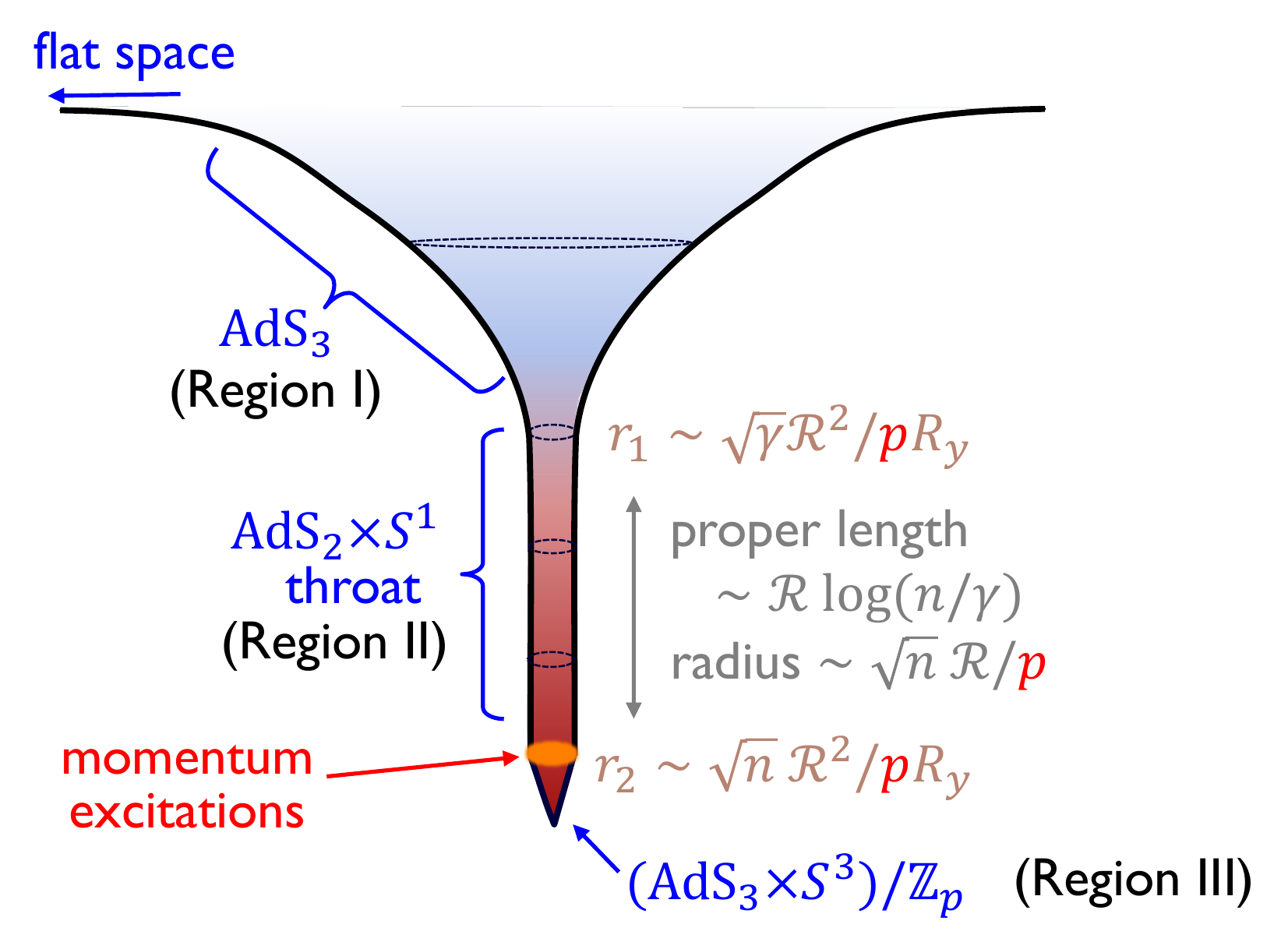} \caption{\sl The schematic
  spacetime structure of the $(1,0,n)_p$ superstratum. The larger $p$
  is, the deeper the throat region starts in the space and the smaller
  the radius of the throat becomes.  The length of the throat is
  independent of $p$.  \label{fig:spacetime_(1,0,n)_p}}
\end{center}
  \end{quote} 
\end{figure}%
We see that, as we go in, we have AdS$_3$, AdS$_2\times S^1$, and global
AdS$_3$ (with orbifolding); see Figure \ref{fig:spacetime_(1,0,n)_p}.
The AdS radius (which is read off from $G_{rr}$) is $\cR$ for all
regions.

The proper radius of the $y$ circle
in the throat region (region II) is
\begin{align}
 \sqrt{G_{yy}}R_y\approx 
{\sqrt{n}\cR\over p}  \qquad (a\ll r \ll b).
\end{align}
So, the throat gets narrower for larger $p$.  The crossover between
regions I and II is at
\begin{align}
 r_1\sim a\sim b\sqrt{\gamma}\sim {\sqrt{\gamma}\,\cR^2\over pR_y},
\end{align}
while the crossover between regions II and III occurs at
\begin{align}
 r_2\sim \sqrt{n}b\sim {\sqrt{n}\,\cR^2\over pR_y}.
\end{align}
So, the position of the throat gets deeper for larger $p$.  On the other
hand, the proper depth of the throat region (region II) along $r$ is
\begin{align}
 \int ds=\int_{r_2}^{r_1} \sqrt{Q_1 Q_5}\Lambda{dr\over\sqrt{r^2+a^2}}
\sim \cR \int_{r_2}^{r_1} {dr\over r}
\sim \cR \log{n\over \gamma},
\end{align}
independent of $p$.
See Figure \ref{fig:spacetime_(1,0,n)_p}.

Independent of $a,b$, we can see that we have a conical singularity at
$r=0,\theta=\pi/2$.  Near this point, we have
\begin{align}
 F_0,F_1\to 1,\qquad F_2\to a^2(2a^2+b^2),\qquad \Lambda\to 1
\end{align}
and therefore the  metric becomes
\begin{equation}
\begin{split}
ds_6^2 \approx\,& \frac{1}{\sqrt{Q_1 Q_5}} \biggl[- \frac{2a^4}{2a^2+ b^2} \;\! dt^2
+ \frac{ r^2 ( 2a^2+b^2 ) }
{2a^2  }
\left( dy + \frac{b^2}{2a^2+b^2} dt \right)^2~\biggr]
\\
&+ \sqrt{Q_1 Q_5}\biggl[  \frac{dr^2}{a^2} + d\theta^2  
+  \sin^2\theta \left( d\phi - 
 \frac{2a^2}{2a^2+b^2} \frac{dt}{p R_y} \right)^2 
\\
& \qquad\qquad\qquad\qquad
 +    \cos^2\theta
\left( d\psi - {dy\over p R_y} -\frac{b^2}{2a^2+b^2}{dt\over pR_y} \right)^2
~\biggr].
\end{split}
\end{equation}
If we define 
\begin{align}
\begin{gathered}
  \yt= {1\over pR_y}\left(y + \frac{b^2}{2a^2+b^2} t\right),\\
\phit= \phi - \frac{2a^2}{2a^2+b^2} \frac{t}{p R_y},\qquad\quad
\psit= \psi - {1\over pR_y}\left(y+\frac{b^2}{2a^2+b^2}t\right),
\end{gathered}\label{ndlg16Dec22}
 \end{align}
the metric becomes
\begin{equation}
\begin{split}
ds_6^2 \approx\,& 
-\frac{1}{\sqrt{Q_1 Q_5}}  \frac{2a^4}{2a^2+ b^2} \;\! dt^2
+
{\sqrt{Q_1 Q_5}\over a^2}  (r^2  d\yt^2+ dr^2)
\\
&\qquad
 + \sqrt{Q_1 Q_5}\left(  d\theta^2  
+  \sin^2\theta\,  d\phit^2  +    \cos^2\theta\, d\psit^2 \right).
\end{split}
\end{equation}
The coordinate identifications are
\begin{align}
 (\phit,\psit,\yt)
 \cong(\phit+2\pi,\psit,\yt)
 \cong(\phit,\psit+2\pi,\yt)
 \cong\left(\phit,\psit-{2\pi\over p},\yt+{2\pi\over p} \right).
\end{align}
So, again, we have the same $\bbZ_p$ singularity.

Note that the coordinate transformation \eqref{ndlg16Dec22} is different
from the spectral flow \eqref{jfjx15Dec22} we used for the empty
$\AdSxS/\bbZ_p$ background or for the $(k,0,0)_p$ superstratum. The
difference vanishes if we take $b=0$. This means that the coordinate
transformation implementing the spectral flow is different near the AdS
boundary and at the bottom of the throat because of redshift, but we do
not have a clear physical understanding of this fact.

\section{Including fractional spectral flows}
\label{sec:fract_spectral_flow}

In the above, we discussed supergravitons/superstrata around the
``background state'' $(\ket{++}_p)^{N/p}$, whose bulk dual is the
conical deficit geometry $\AdSxS/\bbZ_p$.  In \cite{Giusto:2012yz}, the
fractional spectral flow\footnote{In the ``active'' sense in section
\ref{ss:spectral_flow}.} of this state was studied and its bulk dual was
identified.  The dual geometry (the GLMT solution) generically contains
two orbifold singularities.  Here we discuss supergravitons/superstrata
around the GLMT solution.

The CFT state in the R sector representing $\AdSxS/\bbZ_p$ is the R
ground state\footnote{More precisely, it is the Lunin-Mathur geometry
\eqref{LM_geom_p-wound} written in $\phi,\psi$ coordinates that is dual
to this R ground state. $\AdSxS/\bbZ_p$ in \eqref{AdS3xS3/Zp_metric} is
written in terms of $\phit,\psit$ and is in the NS$'$ sector.}
\begin{align}
 (\ket{++}_p)^{N\over p},\qquad h_{\rm R}=\hb_{\rm R}={N\over 4},\qquad 
 j_{\rm R}=\jb_{\rm R}={N\over 2p}.
\end{align}
If we do an active left-spectral flow transformation of this state by a
fractional amount $\eta={s\over p}$, $s\in\bbZ$ (and $\etab=0$), we get
a state which we denote by
\begin{align}
 (\ket{++}_{p,[{s\over p}]})^{N\over p},
\label{fracspecfloweddeficit}
\end{align}
displaying the flow parameter in $[\,]$.  We regard this as a state in
the R sector.   The individual strand has 
the following left-moving quantum numbers:
\begin{align}
 \ket{++}_{p,[{s\over p}]}\;,\qquad
 h_{\rm R}={p\over 4}+{s(s+1)\over p},
 \qquad 
 j_{\rm R}={1\over 2}+s.
\label{fracspecflowindivi}
\end{align}
This state is obtained by filling up the Fermi
surface as follows:
\begin{align}
\ket{++}_{p,[{s\over p}]} &
 =  (\psi^{+\dot{1}}_{-{s-1\over p}}\psi^{+\dot{2}}_{-{s-1\over p}})
 \cdots
 (\psi^{+\dot{1}}_{-{2\over p}}\psi^{+\dot{2}}_{-{2\over p}})
 (\psi^{+\dot{1}}_{-{1\over p}}\psi^{+\dot{2}}_{-{1\over p}})
 \ket{++}_p,
\end{align}
where $\psi^{\alpha \dot{A}}(z)$ is the left-moving fermion with modes
$\psi^{\alpha \dot{A}}_n$.  This expression is valid for $s>0$; for
$s<0$ we instead excite $\psi^{-\dot{A}}$.

Let us denote the bulk GLMT geometry dual to
\eqref{fracspecfloweddeficit} by
\begin{align}
 (\AdSxS/\bbZ_p)_{[{s\over p}]}.
\end{align}
This is nothing but the Lunin-Mathur geometry \eqref{LM_geom_p-wound}
transformed by the bulk spectral flow~\eqref{ieyi18Dec22} with
$\eta={s\over p}$, $\etab=0$.
The strand \eqref{fracspecflowindivi} is not an allowed state of the
symmetric orbifold CFT unless it satisfies the momentum quantization constraint
\begin{align}
 {s(s+1)\over p}\in\bbZ.\label{giuq19Dec22}
\end{align} 
Likewise, the $(\AdSxS/\bbZ_p)_{[{s\over p}]}$
geometry is not physical unless the same quantization condition is
obeyed, as was shown in \cite{Giusto:2012yz} by studying the structure
of orbifold singularities.  Because of the modified coordinate
identifications introduced by the transformation~\eqref{ieyi18Dec22},
this geometry has orbifold singularities at $(r,\theta)=(0,{\pi\over 2})$ and 
$(r,\theta)=(0,0)$ \cite{Giusto:2012yz}.


\bigskip
We can consider the fractional spectral flow of other RR ground states,
such as
\begin{align}
\begin{split}
  \ket{00}_{kp}
 &={1\over\sqrt{2}}\epsilon_{\dot{A}\dot{B}}\,\psi^{-\dot{A}}_0\psit^{-\dot{B}}_0 \ket{++}_{kp}
 ={1\over\sqrt{2}}\epsilon_{\dot{A}\dot{B}}\,\psi^{+\dot{A}}_0\psit^{+\dot{B}}_0 \ket{--}_{kp}
 ,\\
  h_{\rm R}&=\hb_{\rm R}={kp\over 4},\qquad j_{\rm R}=\jb_{\rm R}=0.
\end{split}
\end{align}
If we do a fractional spectral flow of this state by $\eta={s\over p}$,
we get
\begin{align}
 \ket{00}_{kp,[{s\over p}]},\qquad
 h_{\rm R}&={kp\over 4}+{ks^2\over p},\qquad  j_{\rm R}=ks.
\end{align}
This state can be obtained by filling up  each Fermi
surface by $(k s-1)$ levels as:
 \begin{align}
 \ket{00}_{kp,[{s\over p}]}
  =  
 {1\over\sqrt{2}}\epsilon_{\dot{A}\dot{B}}\,
  \psi^{+\dot{A}}_{-{ks\over kp}}\,
 (\psi^{+\dot{1}}_{-{ks-1\over kp}}\psi^{+\dot{2}}_{-{ks-1\over kp}})
 \cdots
 (\psi^{+\dot{1}}_{-{1\over kp}}\psi^{+\dot{2}}_{-{1\over kp}})
  \,
   \psit^{+\dot{B}}_0\,
 \ket{--}_{kp}.
\end{align}
This is an allowed state of the symmetric orbifold CFT if ${ks^2\over p}\in\bbZ$.

More generally, we can take the fractional flow of the $(k,0,0)_p$
superstratum \eqref{ftwj19Dec22} to get
\begin{align}
 (\ket{++}_{p,[{s\over p}]})^{N_0} (\ket{00}_{kp,[{s\over p}]})^{N^{00}_{kp}} ,\qquad
 N_0 + kN^{00}_{kp}={N\over p}.
\end{align}
We can get the bulk dual by taking the $(k,0,0)_p$ geometry of
section \ref{ss:(k,0,0)_p} and applying the bulk spectral
transformation \eqref{ieyi18Dec22} with $\eta={s\over p}$, $\etab=0$.
We can call this a $(k,0,0)_{p,[{s\over p}]}$ superstratum.

\bigskip
In the previous sections, we considered R states such as
\begin{align}
 (\Lt_{-{1\over p}}+\tfrac{1}{p}\Jt^3_{-{1\over p}})^n
 (\Jt^+_{-{1\over p}})^m\ket{00}_{kp},\qquad
 h_{\rm R}={kp\over 4}+{n+m\over p},\quad
 j_{\rm R}=m,\label{igvm18Dec22}
\end{align}
which are dual to the $(k,m,n)_p$ superstratum around $\AdSxS/\bbZ_p$.
If we fractional spectral flow these states, they will become
superstratum states around $(AdS_3\times S^3/\bbZ_p)_{[{s\over p}]}$,
which is obtained by taking known superstratum backgrounds
\cite{Bena:2015bea, Bena:2016ypk, Bena:2017xbt}, setting $R_y\to p R_y$,
and applying the bulk spectral transformation \eqref{ieyi18Dec22} with
$\eta={s\over p}$, $\etab=0$.

Under spectral flow, the generators get transformed as
\begin{align}
 L_{n}\to L_n,\qquad
 J^\pm_{n}\to J^\pm_{n\mp 2\eta},\qquad
 J^3_{n}\to J^3_n,\qquad
 G^{\pm,A}_{n} \to G^{\pm,A}_{n\mp \eta }.
\end{align}
Therefore, for example, the state \eqref{igvm18Dec22} goes to
\begin{align}
 (\Lt_{-{1\over p}}+\tfrac{1}{p}\Jt^3_{-{1\over p}})^n
 (\Jt^+_{-{1+2s\over p}})^m\ket{00}_{kp,[{s\over p}]}\label{meen26Dec22}
\end{align}
with
\begin{align}
 h_{\rm R}^{\rm stratum}={kp\over 4}+{ks^2\over p}+{n+m+2ms\over p},\qquad
 j_{\rm R}=m+ks.\label{lkve18Dec22}
\end{align}
This is an allowed state of the CFT if
\begin{align}
 {ks^2\over p}+{n+m+2ms\over p}\in\bbZ.\label{gnoj19Dec22}
\end{align}

Let us look at the bulk side.  The bulk mode \eqref{jhmr18Dec22} on
$\AdSxS/\bbZ_p$ contains the phase~\eqref{jicf18Dec22}.  From
\eqref{ieyi18Dec22}, the bulk spectral transformation with $\eta={s\over
p},\etab=0$ amounts to the replacement
\begin{align}
 \phi &\to \phi  - {s\over p} {t+y\over R_y},\qquad
 \psi  \to \psi  - {s\over p} {t+y\over R_y}.
\end{align}
So, the phase of the bulk solution is
\begin{align}
e^{iv_{kmn}'}
 \to \exp\Bigl[i\Bigl(
(m+n+(2m-k)s){t+y\over p R_y}+(k-m)\phi -m\psi\Bigr)
\Bigr].\label{ngkm18Dec22}
\end{align}
One can ask two questions.  First, why is the energy read off from this,
$h^{\rm bulk}={m+n+(2m-k)s\over p}$, different from the CFT expression
\eqref{lkve18Dec22}? For this, note that $h^{\rm bulk}$ is the energy of
the propagating wave in the $(\AdSxS/\bbZ_p)_{[{s\over p}]}$ background;
namely, it does not contain the energy carried by the background itself.
The background state has, from \eqref{fracspecflowindivi},
\begin{align}
 (\ket{++}_{p,[{s\over p}]})^k,\qquad
 h_{\rm R}^{\rm backgnd}=k\cdot\left({p\over 4}+{s(s+1)\over p}\right).\label{nhed18Dec22}
\end{align}
So, the energy to be compared with $h^{\rm bulk}$ is
\begin{align}
 h_{\rm R}^{\rm stratum}- h_{\rm R}^{\rm backgnd}
 ={n+m+(2m-k)s\over p}
\end{align}
which is equal to $h^{\rm bulk}$.  The second thing is that requirement we
get from the single-valuedness of the phase factor,
\begin{align}
 {m+n+(2m-k)s\over p}\in\bbZ,
\end{align}
looks different from the CFT condition, \eqref{gnoj19Dec22}.  However,
recalling that we must satisfy the quantization condition for the
background, \eqref{giuq19Dec22}, we can immediately show that the two
conditions are equivalent.

\bigskip
The fractional spectral flowed state 
\eqref{fracspecfloweddeficit}
dual to the bulk background geometry\linebreak
$(\AdSxS/\bbZ_p)_{[{s\over p}]}$ is not a valid state unless the
quantization condition \eqref{giuq19Dec22} is met. If we in addition
excite strands \eqref{meen26Dec22}, we need to simultaneously satisfy
the quantization condition \eqref{gnoj19Dec22}.  However, if we excite
as many strands of this kind \eqref{meen26Dec22} as possible, namely, if
the total state is
\begin{align}
 \bigl((\Lt_{-{1\over p}}+\tfrac{1}{p}\Jt^3_{-{1\over p}})^n
 (\Jt^+_{-{1+2s\over p}})^m\ket{00}_{kp,[{s\over p}]}\bigr)^{N\over kp}
\end{align}
then we only have to satisfy \eqref{gnoj19Dec22}. This points toward an
interesting possibility -- even if the background is not allowed, if we
excite the maximum possible number of certain strands, the state becomes
allowed.  This kind of states may lead to a larger
entropy.\footnote{This may be related to the fact that, for $\cM=T^4$,
the only states that contribute to the modified elliptic genus are the
ones with identical strands with the same states on them
\cite{Maldacena:1999bp}.}  In such a limit, the six-dimensional metric
becomes indistinguishable from the black hole geometry (this can be
seen, e.g., in the $(1,0,n)$ superstratum by setting $a\to 0$).  If we
want to see the structure of such states we must probably look at the
structure in the internal manifold $\cM$.  This may be suggesting that
much of black-hole microstructure lies inside the internal manifold
\cite{Maldacena:1997de, Bena:2022fzf}.


\section{Discussion}
\label{sec:disc} 

In this paper, we gave a CFT perspective on the construction of the
superstrata on the orbifold background $\AdSxS/\bbZ_p$ and provided some
analysis of the explicit geometry in the bulk.  We also discussed
generalization to superstrata on the fractional spectral flowed
background $(\AdSxS/\bbZ_p)_{[{s\over p}]}$.

Given these generalizations of superstratum solutions, which is known to
have a large entropy, one immediate question is what the entropy of
these new solutions is.  In particular, because these generalized
superstrata involve fractional modes that have been argued to be
crucial for reproducing the Strominger-Vafa entropy, one naturally
wonders whether they reproduce the Strominger-Vafa entropy.  The answer
turns out to be negative.

The problem of counting superstrata is roughly (ignoring angular
momentum and degeneracies coming from different species of anti-chiral
primaries based on which these solutions are built) the problem of
counting $\{N_{k,m,n}\}$ satisfying
\begin{align}
 \sum_{k,m,n}kN_{k,m,n}=N,\qquad
 \sum_{k,m,n}(m+n)N_{k,m,n}=N_P.
\end{align}
The resulting entropy is \cite{Shigemori:2019orj, Mayerson:2020acj}
\begin{align}
 S\sim N^{1/2}N_P^{1/4}.\label{gxjm19Dec22}
\end{align}
The entropy for superstrata on $\AdSxS/\bbZ_p$ can roughly be evaluated
by counting $\{N_{k,m,n}\}$ satisfying
\begin{align}
 \sum_{k,m,n}kpN_{k,m,n}=N,\qquad
 \sum_{k,m,n}{m+n\over p}N_{k,m,n}=N_P.\label{figm22Dec22}
\end{align}
The first one is because now all strand lengths are integer multiples of
$p$, and the second one is because all momenta are now $1\over
p$-moded.  By moving $p$ in \eqref{figm22Dec22}
 to the right-hand side, one sees that this
changes the entropy \eqref{gxjm19Dec22} to
\begin{align}
 S\sim \left({N\over p}\right)^{1/2}(p N_P)^{1/4} 
 =p^{-1/4}N^{1/2}N_P^{1/4}.\label{fifw22Dec22}
\end{align}
Namely, fractionation reduces entropy.  Having fractional momenta
enhances entropy, but that is more than canceled by the strand-length
budget getting tighter.  To be more precise, the sum in
\eqref{figm22Dec22} must be restricted with the condition ${m+n\over
p}\in \bbZ$, but that is a minor condition for large $N,N_P$ and should
not change the estimate \eqref{fifw22Dec22}.

One can also consider the fractional spectral flow of superstrata, or
superstrata on GLMT solutions, discussed in section
\ref{sec:fract_spectral_flow}.  We have not estimated their entropy, one
technical issue being that states in different $(p,[{s\over p}])$
sectors are not necessarily independent.  However, but one extra parameter
$(s)$ is very unlikely to lead to a parametrically larger entropy.

\bigskip
The fact that superstrata on $\AdSxS/\bbZ_p$ and GLMT backgrounds do not
have the Strominger-Vafa entropy has interesting implications.  In
\cite{Bossard:2019ajg}, it was shown that the supersymmetric microstate
geometries in six dimensions that contribute to supersymmetry indices
must have only one compact 3-cycle.  However, $\AdSxS/\bbZ_p$ and GLMT
backgrounds are the most general 3-cycles that can be written as
bubbling multi-center solutions.  Therefore, unless there are more
exotic possibilities for 3-cycles in six dimensions and excitations on
them, the generalized superstrata considered in the current paper are
the most general microstate geometries that are counted by supersymmetry
indices.  The fact that they do not reproduce the Strominger-Vafa
entropy likely means that six-dimensional microstate geometries are
simply not enough for reproducing the entropy.  One either have to
consider stringy degrees of freedom (see \cite{Martinec:2017ztd,
Martinec:2018nco, Martinec:2019wzw, Martinec:2020gkv, Bufalini:2021ndn,
Bufalini:2022wyp, Bufalini:2022wzu, Martinec:2022okx} for recent
developments in the worldsheet theory for the relevant systems), or
remain in supergravity but look for microstate geometries in $d>6$
dimensions, such as backreacted MSW M5-brane waves
\cite{Maldacena:1997de}, or more recently proposed super-maze
\cite{Bena:2022wpl} and hyperstrata \cite{Bena:2022fzf}.

Based on the above argument, one naturally expects that generalized
superstrata considered in the current paper are the most general
supergravity solutions that contribute to supersymmetry indices.  In
\cite{deBoer:1998us} (see also \cite{Maldacena:1999bp}), it was shown
that the K3 elliptic genus for $L_0^{\rm NS}\le {N+1\over 4}$ can be
completely accounted for by supergravitons, i.e., superstrata.  It is
possible that, above this bound and below the black-hole bound,
$L_0^{\rm NS}= {J^2\over 4N}+{N\over 4}$ where new states come in, the
elliptic genus can be accounted for by counting the generalized
superstrata.  It would be quite interesting to check if this is correct
or not.  One technical obstacle is that states in different $(p,[{s\over
p}])$ sectors are not necessarily independent.  Counting supergravitons
for different values of $p,s$ separately would be overcounting.  One
needs to develop technical tools to properly count the supergraviton
states on those backgrounds.


\section*{Acknowledgments}

I would like to thank Emil Martinec and Nick Warner for discussions.  I
would like to thank CEA Saclay for their hospitality in the ``Black-Hole
Microstructure IV'' workshop.  This work was supported in part by MEXT
KAKENHI Grant Numbers 21H05184 and 21K03552.

\appendix

\section{Supergravity solutions}
\label{app:sugra_soln}

Here we summarize the supergravity fields that are used in the main
text.

The solution preserving the same supersymmetry as the D1-D5-P black hole
and preserves the symmetry of the internal manifold $\cM$ ($=T^4$ or K3)
has the following 10-dimensional fields \cite[Appendix
E]{Giusto:2013rxa}:
\begin{subequations}\label{ansatzSummary}
 \begin{align}
d s^2_{10,str} & = \sqrt{Z_1 Z_2\over \cP}\,ds_6^2 + \sqrt{\frac{Z_1}{Z_2}}\,ds^2(\cM),\label{10dmetric}\\
ds_{6E}^2  &=-\frac{2}{\sqrt{\cP}}(d v+\beta)\Big[d u+\omega + \frac{\mathcal{F}}{2}(d v+\beta)\Big]+\sqrt{\cP}\,d s^2(\cB),\\
e^{2\Phi}&={Z_1^2\over \cP} ,\qquad
B_2= -\frac{Z_4}{\cP}(d u+\omega) \wedge(d v+\beta)+ a_4 \wedge  (d v+\beta) + \delta_2, \label{Bform}\\ 
C_0&=\frac{Z_4}{Z_1} ,\qquad
C_2 = -{Z_2 \over \cP}(d u+\omega) \wedge(d v+\beta)+ a_1 \wedge  (d v+\beta) + \gamma_2,\\ 
C_4 &= \frac{Z_4}{Z_2} \mathrm{vol}(\cM) - \frac{Z_4}{\cP}\gamma_2\wedge (d u+\omega) \wedge(d v+\beta)+x_3\wedge(d v + \beta) ,\\
C_6 &=\mathrm{vol}(\cM) \wedge \left[ -{Z_1\over \cP}(d u+\omega) \wedge(d v+\beta)+ a_2 \wedge  (d v+\beta) + \gamma_1\right] 
\label{jtta10Apr18} 
\end{align}
\end{subequations}
where
\begin{align}
\cP \equiv Z_1\,Z_2 - Z_4^2.
\end{align}
For equations to be satisfied by various ansatz quantities appearing
here, as the scalars $Z_{1,2,4},\cF$, the 1-forms
$\omega,\beta,a_{1,2,4}$, the 2-forms $\gamma_{1,2},\delta$, and the
3-form $x_3$, see \cite[Appendix E]{Giusto:2013rxa}.  The coordinates
$u,v$ are related to the time coordinate $t$ and the coordinate $y$ of
the $S^1$ with periodicity $2\pi R_y$ as
\begin{align}\label{sptvw}
u = \frac{1}{\sqrt{2}}(t-y), \qquad v = \frac{1}{\sqrt{2}}(t+y).
\end{align}
The field strengths can be written in terms of the gauge invariant 
combinations
\begin{gather}
 \begin{aligned}
 \Theta_1&\equiv \cD a_1+\dot{\gamma}_2-\dot{\beta}\wedge a_1,~&
 \Theta_2&\equiv \cD a_2+\dot{\gamma}_1-\dot{\beta}\wedge a_2,~&
 \Theta_4&\equiv \cD a_4+\dot{\delta}_2-\dot{\beta}\wedge a_4,
 \end{aligned}
\end{gather}
where $\cD\equiv d_4-\beta\wedge \p_v$ and $\dot{~}\equiv\p_v$.  $d_4$ is the
exterior derivative in the four-dimensional base space $\cB$. This base
space $\cB$ has metric $d s^2(\cB)$ and can generally be an almost
hyper-K\"ahler space, but in this paper we take it to be flat $\bbR^4$
with metric
\begin{subequations} 
 \label{def_base_metric_flat_Sigma}
 \begin{align}
 \label{base_metric_flat}
 d s^2(\cB) &= \Sigma\, \Bigl(\frac{d r^2}{r^2+a^2}+ d\theta^2\Bigr)+(r^2+a^2)\sin^2\theta\,d\phi^2+r^2 \cos^2\theta\,d\psi^2,\\
 \Sigma&\equiv r^2+a^2 \cos^2\theta
 .\label{eqmm19Dec22}
 \end{align}
\end{subequations}
The relation to the Cartesian coordinates $x^i$ of $\bbR^4$ is
\begin{align}
 x^1+ix^2=\sqrt{r^2+a^2}\sin\theta\,e^{i\phi},\qquad
 x^3+ix^4=r\cos\theta\,e^{i\psi}.
\end{align}

We use the following definitions in the main text:
\begin{subequations} 
 \label{Delta_v_kmn_def}
 \begin{align}
  \Delta_{k,m,n} &\equiv
 \left(\frac{a}{\sqrt{r^2+a^2}}\right)^k
 \left(\frac{r}{\sqrt{r^2+a^2}}\right)^n 
 \cos^{m}\theta \, \sin^{k-m}\theta , 
 \label{Delta_kmn_def}
 \\
 v_{k,m,n} &\equiv (m+n) \frac{\sqrt{2}\,v}{R_y} + (k-m)\phi - m\psi ,
 \label{v_kmn_def}
 \end{align}%
\end{subequations}%
\begin{align}
 \vartheta_{k,m,n}&\equiv -\sqrt{2}\,
 \Delta_{k,m,n}
 \biggl[\biggl((m+n)\,r\sin\theta +n\left({m\over k}-1\right){\Sigma\over r \sin\theta}  \biggr)\Omega^{(1)}\sin{v_{k,m,n}} \nonumber\\
 &\hspace{16ex}
 +\biggl(m\left({n\over k}+1\right)\Omega^{(2)} +\left({m\over k}-1\right)n\, \Omega^{(3)}\biggr) \cos{v_{k,m,n}} \biggr]
    \label{theta_kmn}
 \,,
\\
\label{selfdualbasis}
&
\begin{aligned}
\Omega^{(1)} &\equiv \frac{dr\wedge d\theta}{(r^2+a^2)\cos\theta} + \frac{r\sin\theta}{\Sigma} d\phi\wedge d\psi\,,\\
\Omega^{(2)} &\equiv  \frac{r}{r^2+a^2} dr\wedge d\psi + \tan\theta\, d\theta\wedge d\phi\,,\\
 \Omega^{(3)}&\equiv \frac{dr\wedge d\phi}{r} - \cot\theta\, d\theta\wedge d\psi\,.
\end{aligned}
\end{align}

\subsection{Lunin-Mathur geometries}

The Lunin-Mathur geometries 
\cite{Lunin:2001jy, Lunin:2002iz, Taylor:2005db,
Kanitscheider:2007wq} that
respect the symmetry of the internal manifold $\cM$ are parametrized by
profile functions $g_A(\lambda)$, $A=1,2,3,4,5$ with periodicity $L$.
The ansatz data are given by
\begin{subequations}\label{LM_geom_general}
\begin{align}
 Z_1 &= 1 + \frac{Q_5}{L} \int_0^{L} d\lambda \frac{|\partial_\lambda g_i(\lambda )|^2+|\partial_\lambda g_5(\lambda )|^2}{|x_i -g_i(\lambda )|^2} ,\qquad
  Z_4 = - \frac{Q_5}{L} \int_0^{L} d\lambda \frac{\partial_\lambda g_5(\lambda )}{|x_i -g_i(\lambda )|^2} ,\\\label{Z1profile}
 Z_2 &= 1 + \frac{Q_5}{L} \int_0^{L} \frac{d\lambda }{|x_i -g_i(\lambda )|^2}, \qquad d\gamma_2 = *_4 d Z_2,\qquad d\delta_2 = *_4 d Z_4,\\
 A &= - \frac{Q_5}{L} dx^j \int_0^{L} d\lambda  \frac{\partial_\lambda g_j(\lambda )}{|x_i -g_i(\lambda )|^2} , \qquad dB = - {*_4 dA},\qquad ds^2(\cB) = dx^i dx^i, \label{general2chg-A,B}\\
 \beta &= \frac{-A+B}{\sqrt{2}},\qquad\omega = \frac{-A-B}{\sqrt{2}},\qquad
 \Theta_I=\cF=a_{1,4}=x_3=0,\\
 Q_1&={Q_5\over L}\int_0^L d\lambda |\partial_\lambda g_i(\lambda )|^2+|\partial_\lambda g_5(\lambda )|^2
\end{align}
The 6D part of the metric can be written as
\begin{equation}
 ds_{6E}^2  =\frac{1}{\sqrt{\cP}}\left[-(dt-A)^2+(dy+B)^2\right]+\sqrt{\cP}\,d s^2(\cB).
\end{equation}
\end{subequations}
The supergravity charges $Q_1,Q_5$ are related to the D1 and D5 numbers
$N_1,N_5$ by
\begin{align}
 Q_1={N_1 g_s \alpha'^3\over v_4},\qquad Q_5=N_5 g_s \alpha',
\end{align}
where $(2\pi)^4v_4$ is the coordinate volume of $\cM$.

Below we discuss some explicit solutions for particular profile
functions.  For more details see 
\cite{Shigemori:2020yuo} and references therein.

\subsubsection{$\AdSxS/\bbZ_p$}
\label{app:sss:AdS3xS3/Zp}

As the profile, take a $p$-times wound circle,
\begin{align}
 g_1+ig_2=a e^{2\pi i p\lambda/L},\qquad g_3+ig_4=g_5=0,\label{profile_emptyAdS3xS3/Zp}
 \qquad p\ge 1,
\end{align}
whose
CFT dual in the R sector is
\begin{align}
  [\ket{++}_p]^{N/p}.\label{bfcg23Jan20}
\end{align}
The ansatz data are
\begin{subequations}
 \label{ansatz_data_empty_AdS3xS3/Zp}
 \begin{align}
 Z_1&=1+{(p R_y)^2a^2\over Q_5\Sigma},\qquad
 Z_2=1+{Q_5\over \Sigma},\qquad
 Z_4=0,\qquad \Theta_I=0,
 \\
 \beta&=  \frac{p R_y a^2}{\sqrt{2}\,\Sigma}(\sin^2\theta\, d\phi - \cos^2\theta\,d\psi)= p\beta_0,
 \\
 \omega&= \frac{p R_y a^2}{\sqrt{2}\,\Sigma}(\sin^2\theta\, d\phi + \cos^2\theta\,d\psi)=p \omega_0,\qquad \cF=0,
 \end{align}
\end{subequations}
and
\begin{align}
 a^2={Q_1 Q_5\over (p R_y)^2}.\label{aQR_y/Zp}
\end{align}
One sees that the only effect of introducing $p$ is to replace $R_y\to p R_y$.

The decoupling limit amounts to dropping ``1'' in $Z_{1,2}$, after which
we obtain \eqref{LM_geom_p-wound}.

\subsubsection{$(k,0,0)_p$ superstratum}
\label{app:sss:(k,0,0)_p}

Consider a $p$-times wound circle, with $g_5$ turned on:
\begin{align}
 g_1+ig_2=ae^{2\pi i p\lambda /L},\qquad g_3+ig_4=0,\qquad
 g_5=-{b\over k}\sin{2\pi   k p\lambda \over L}
\label{circularLM+g5}
\end{align}
where $k\in\bbZ$.
This is  dual to the CFT state
\begin{align}
 (\ket{++}_p)^{N_0} (\ket{00}_{kp})^{N^{00}_{kp}} ,\qquad
 N_0 + kN^{00}_{kp}={N\over p}
\end{align}
where $N_0\propto a^2,N_k^{00}\propto b^2$.
We can call this a $(k,0,0)_p$ superstratum.

The supergravity ansatz data are 
\begin{align}
 \label{Z4_excitaion_ansatz_data_p}
\begin{split}
 Z_1 &= 1+\frac{(pR_y)^2}{Q_5}\Bigl[\frac{a^2+{b^2/ 2}}{\Sigma} 
 + b^2 a^{2k} \frac{\sin^{2k}\theta\,\cos(2k\phi)}{2(r^2+a^2)^k\,\Sigma}\Bigr],
 \qquad Z_2 = 1+\frac{Q_5}{\Sigma}\,, \\
 Z_4&=(pR_y) b a^k \frac{\sin^k\theta\,\cos(k\phi)}{(r^2+a^2)^{k/2}\,\Sigma},\qquad
 \cF=\Theta_I=0,\\
 \beta&=  \frac{(p R_y) a^2}{\sqrt{2}\,\Sigma}(\sin^2\theta\, d\phi - \cos^2\theta\,d\psi)=p \beta_0,\\
 \omega&= \frac{(p R_y) a^2}{\sqrt{2}\,\Sigma}(\sin^2\theta\, d\phi + \cos^2\theta\,d\psi)=p \omega_0,
\end{split}
\end{align}
where $a$ and $b$ must satisfy the budget relation
\begin{align}
 a^2+{b^2\over 2}={Q_1 Q_5\over (pR_y)^2}.\label{egjh30Jun22}
\end{align}
One sees that the only effect of introducing $p$ is to replace $R_y\to p R_y$.



\begin{thebibliography}{99}

\bibitem{Bena:2013dka}
I.~Bena and N.~P.~Warner,
``Resolving the Structure of Black Holes: Philosophizing with a Hammer,''
[arXiv:1311.4538 [hep-th]].

\bibitem{Bena:2022ldq}
I.~Bena, E.~J.~Martinec, S.~D.~Mathur and N.~P.~Warner,
``Snowmass White Paper: Micro- and Macro-Structure of Black Holes,''
[arXiv:2203.04981 [hep-th]].

\bibitem{Bena:2022rna}
I.~Bena, E.~J.~Martinec, S.~D.~Mathur and N.~P.~Warner,
``Fuzzballs and Microstate Geometries: Black-Hole Structure in String Theory,''
[arXiv:2204.13113 [hep-th]].


\bibitem{Strominger:1996sh}
A.~Strominger and C.~Vafa,
``Microscopic origin of the Bekenstein-Hawking entropy,''
Phys. Lett. B \textbf{379}, 99-104 (1996)
[arXiv:hep-th/9601029 [hep-th]].


\bibitem{Bena:2004de}
I.~Bena and N.~P.~Warner,
``One ring to rule them all ... and in the darkness bind them?,''
Adv. Theor. Math. Phys. \textbf{9}, no.5, 667-701 (2005)
[arXiv:hep-th/0408106 [hep-th]].

\bibitem{Gauntlett:2004qy}
J.~P.~Gauntlett and J.~B.~Gutowski,
``General concentric black rings,''
Phys. Rev. D \textbf{71}, 045002 (2005)
[arXiv:hep-th/0408122 [hep-th]].

\bibitem{Bena:2005va}
I.~Bena and N.~P.~Warner,
``Bubbling supertubes and foaming black holes,''
Phys. Rev. D \textbf{74}, 066001 (2006)
[arXiv:hep-th/0505166 [hep-th]].

\bibitem{Berglund:2005vb}
P.~Berglund, E.~G.~Gimon and T.~S.~Levi,
``Supergravity microstates for BPS black holes and black rings,''
JHEP \textbf{06}, 007 (2006)
[arXiv:hep-th/0505167 [hep-th]].

\bibitem{Bena:2006kb}
I.~Bena, C.~W.~Wang and N.~P.~Warner,
``Mergers and typical black hole microstates,''
JHEP \textbf{11}, 042 (2006)
[arXiv:hep-th/0608217 [hep-th]].

\bibitem{Bena:2007kg}
I.~Bena and N.~P.~Warner,
``Black holes, black rings and their microstates,''
Lect. Notes Phys. \textbf{755}, 1-92 (2008)
[arXiv:hep-th/0701216 [hep-th]].

\bibitem{Bena:2007qc}
I.~Bena, C.~W.~Wang and N.~P.~Warner,
``Plumbing the Abyss: Black ring microstates,''
JHEP \textbf{07}, 019 (2008)
[arXiv:0706.3786 [hep-th]].

\bibitem{Bena:2010gg}
I.~Bena, N.~Bobev, S.~Giusto, C.~Ruef and N.~P.~Warner,
``An Infinite-Dimensional Family of Black-Hole Microstate Geometries,''
JHEP \textbf{03}, 022 (2011)
[erratum: JHEP \textbf{04}, 059 (2011)]
[arXiv:1006.3497 [hep-th]].

\bibitem{Bianchi:2017bxl}
M.~Bianchi, J.~F.~Morales, L.~Pieri and N.~Zinnato,
``More on microstate geometries of 4d black holes,''
JHEP \textbf{05}, 147 (2017)
[arXiv:1701.05520 [hep-th]].

\bibitem{Heidmann:2017cxt}
P.~Heidmann,
``Four-center bubbled BPS solutions with a Gibbons-Hawking base,''
JHEP \textbf{10}, 009 (2017)
[arXiv:1703.10095 [hep-th]].

\bibitem{Bena:2017fvm}
I.~Bena, P.~Heidmann and P.~F.~Ramirez,
``A systematic construction of microstate geometries with low angular momentum,''
JHEP \textbf{10}, 217 (2017)
[arXiv:1709.02812 [hep-th]].

\bibitem{Avila:2017pwi}
J.~Avila, P.~F.~Ramirez and A.~Ruiperez,
``One Thousand and One Bubbles,''
JHEP \textbf{01}, 041 (2018)
[arXiv:1709.03985 [hep-th]].

\bibitem{Mayerson:2022yoc}
D.~R.~Mayerson,
``Modave Lectures on Horizon-Size Microstructure, Fuzzballs and Observations,''
[arXiv:2202.11394 [hep-th]].

\bibitem{Rawash:2022sum}
S.~Rawash and D.~Turton,
``Evolutionary algorithms for multi-center solutions,''
[arXiv:2212.08585 [hep-th]].

\bibitem{Giusto:2004id}
S.~Giusto, S.~D.~Mathur and A.~Saxena,
``Dual geometries for a set of 3-charge microstates,''
Nucl. Phys. B \textbf{701}, 357-379 (2004)
[arXiv:hep-th/0405017 [hep-th]].

\bibitem{Giusto:2004ip}
S.~Giusto, S.~D.~Mathur and A.~Saxena,
``3-charge geometries and their CFT duals,''
Nucl. Phys. B \textbf{710}, 425-463 (2005)
[arXiv:hep-th/0406103 [hep-th]].

\bibitem{Lunin:2004uu}
O.~Lunin,
``Adding momentum to D-1 - D-5 system,''
JHEP \textbf{04}, 054 (2004)
[arXiv:hep-th/0404006 [hep-th]].

\bibitem{Jejjala:2005yu}
V.~Jejjala, O.~Madden, S.~F.~Ross and G.~Titchener,
``Non-supersymmetric smooth geometries and D1-D5-P bound states,''
Phys. Rev. D \textbf{71}, 124030 (2005)
[arXiv:hep-th/0504181 [hep-th]].


\bibitem{Giusto:2012yz} 
  S.~Giusto, O.~Lunin, S.~D.~Mathur and D.~Turton,
  ``D1-D5-P microstates at the cap,''
  JHEP {\bf 1302}, 050 (2013)
  [arXiv:1211.0306 [hep-th]].

\bibitem{Chakrabarty:2015foa}
B.~Chakrabarty, D.~Turton and A.~Virmani,
``Holographic description of non-supersymmetric orbifolded D1-D5-P solutions,''
JHEP \textbf{11}, 063 (2015)
[arXiv:1508.01231 [hep-th]].

\bibitem{Bena:2015bea}
I.~Bena, S.~Giusto, R.~Russo, M.~Shigemori and N.~P.~Warner,
``Habemus Superstratum! A constructive proof of the existence of superstrata,''
JHEP \textbf{05}, 110 (2015)
[arXiv:1503.01463 [hep-th]].


\bibitem{Bena:2016agb} 
  I.~Bena, E.~Martinec, D.~Turton and N.~P.~Warner,
  ``Momentum Fractionation on Superstrata,''
  JHEP {\bf 1605}, 064 (2016)
  [arXiv:1601.05805 [hep-th]].

\bibitem{Bena:2016ypk}
I.~Bena, S.~Giusto, E.~J.~Martinec, R.~Russo, M.~Shigemori, D.~Turton and N.~P.~Warner,
``Smooth horizonless geometries deep inside the black-hole regime,''
Phys. Rev. Lett. \textbf{117}, no.20, 201601 (2016)
[arXiv:1607.03908 [hep-th]].

\bibitem{Bena:2017geu}
I.~Bena, E.~Martinec, D.~Turton and N.~P.~Warner,
``M-theory Superstrata and the MSW String,''
JHEP \textbf{06}, 137 (2017)
[arXiv:1703.10171 [hep-th]].


\bibitem{Bena:2017xbt} 
  I.~Bena, S.~Giusto, E.~J.~Martinec, R.~Russo, M.~Shigemori, D.~Turton and N.~P.~Warner,
  ``Asymptotically-flat supergravity solutions deep inside the black-hole regime,''
  JHEP {\bf 1802}, 014 (2018)
  [arXiv:1711.10474 [hep-th]].

\bibitem{Ceplak:2018pws}
N.~\v{C}eplak, R.~Russo and M.~Shigemori,
``Supercharging Superstrata,''
JHEP \textbf{03}, 095 (2019)
[arXiv:1812.08761 [hep-th]].

\bibitem{Heidmann:2019zws}
P.~Heidmann and N.~P.~Warner,
``Superstratum Symbiosis,''
JHEP \textbf{09}, 059 (2019)
[arXiv:1903.07631 [hep-th]].

\bibitem{Heidmann:2019xrd}
P.~Heidmann, D.~R.~Mayerson, R.~Walker and N.~P.~Warner,
``Holomorphic Waves of Black Hole Microstructure,''
JHEP \textbf{02}, 192 (2020)
[arXiv:1910.10714 [hep-th]].

\bibitem{Ganchev:2022exf}
B.~Ganchev, A.~Houppe and N.~P.~Warner,
``Elliptical and purely NS superstrata,''
JHEP \textbf{09}, 067 (2022)
[arXiv:2207.04060 [hep-th]].

\bibitem{Ceplak:2022pep}
N.~\v{C}eplak,
``Vector Superstrata,''
[arXiv:2212.06947 [hep-th]].

\bibitem{Shigemori:2020yuo}
M.~Shigemori,
``Superstrata,''
Gen. Rel. Grav. \textbf{52}, no.5, 51 (2020)
[arXiv:2002.01592 [hep-th]].

\bibitem{Maldacena:1998bw}
J.~M.~Maldacena and A.~Strominger,
``AdS(3) black holes and a stringy exclusion principle,''
JHEP \textbf{12}, 005 (1998)
[arXiv:hep-th/9804085 [hep-th]].

\bibitem{Deger:1998nm}
S.~Deger, A.~Kaya, E.~Sezgin and P.~Sundell,
``Spectrum of D = 6, N=4b supergravity on AdS in three-dimensions x S**3,''
Nucl. Phys. B \textbf{536}, 110-140 (1998)
[arXiv:hep-th/9804166 [hep-th]].

\bibitem{Larsen:1998xm}
F.~Larsen,
``The Perturbation spectrum of black holes in N=8 supergravity,''
Nucl. Phys. B \textbf{536}, 258-278 (1998)
[arXiv:hep-th/9805208 [hep-th]].

\bibitem{deBoer:1998kjm}
J.~de Boer,
``Six-dimensional supergravity on S**3 x AdS(3) and 2-D conformal field theory,''
Nucl. Phys. B \textbf{548}, 139-166 (1999)
[arXiv:hep-th/9806104 [hep-th]].

\bibitem{Giusto:2015dfa}
S.~Giusto, E.~Moscato and R.~Russo,
``AdS$_{3}$ holography for 1/4 and 1/8 BPS geometries,''
JHEP \textbf{11}, 004 (2015)
[arXiv:1507.00945 [hep-th]].

\bibitem{Giusto:2019qig}
S.~Giusto, S.~Rawash and D.~Turton,
``Ads$_{3}$ holography at dimension two,''
JHEP \textbf{07}, 171 (2019)
[arXiv:1904.12880 [hep-th]].

\bibitem{Rawash:2021pik}
S.~Rawash and D.~Turton,
``Supercharged AdS$_{3}$ Holography,''
JHEP \textbf{07}, 178 (2021)
[arXiv:2105.13046 [hep-th]].

\bibitem{Ganchev:2021ewa}
B.~Ganchev, S.~Giusto, A.~Houppe and R.~Russo,
``$\hbox {AdS}_3$ holography for non-BPS geometries,''
Eur. Phys. J. C \textbf{82}, no.3, 217 (2022)
[arXiv:2112.03287 [hep-th]].

\bibitem{Guo:2022ifr}
B.~Guo, M.~R.~R.~Hughes, S.~D.~Mathur and M.~Mehta,
``Universal lifting in the D1-D5 CFT,''
JHEP \textbf{10}, 148 (2022)
[arXiv:2208.07409 [hep-th]].

\bibitem{Tyukov:2017uig}
A.~Tyukov, R.~Walker and N.~P.~Warner,
``Tidal Stresses and Energy Gaps in Microstate Geometries,''
JHEP \textbf{02}, 122 (2018)
[arXiv:1710.09006 [hep-th]].

\bibitem{Bena:2018mpb}
I.~Bena, E.~J.~Martinec, R.~Walker and N.~P.~Warner,
``Early Scrambling and Capped BTZ Geometries,''
JHEP \textbf{04}, 126 (2019)
[arXiv:1812.05110 [hep-th]].

\bibitem{Raju:2018xue}
S.~Raju and P.~Shrivastava,
``Critique of the fuzzball program,''
Phys. Rev. D \textbf{99}, no.6, 066009 (2019)
[arXiv:1804.10616 [hep-th]].

\bibitem{Bena:2018bbd}
I.~Bena, P.~Heidmann and D.~Turton,
``AdS$_{2}$ holography: mind the cap,''
JHEP \textbf{12}, 028 (2018)
[arXiv:1806.02834 [hep-th]].

\bibitem{Bianchi:2018kzy}
M.~Bianchi, D.~Consoli, A.~Grillo and J.~F.~Morales,
``The dark side of fuzzball geometries,''
JHEP \textbf{05}, 126 (2019)
[arXiv:1811.02397 [hep-th]].

\bibitem{Bena:2019azk}
I.~Bena, P.~Heidmann, R.~Monten and N.~P.~Warner,
``Thermal Decay without Information Loss in Horizonless Microstate Geometries,''
SciPost Phys. \textbf{7}, no.5, 063 (2019)
[arXiv:1905.05194 [hep-th]].

\bibitem{Bombini:2019vnc}
A.~Bombini and A.~Galliani,
``AdS$_{3}$ four-point functions from $ \frac{1}{8} $ -BPS states,''
JHEP \textbf{06}, 044 (2019)
[arXiv:1904.02656 [hep-th]].

\bibitem{Tian:2019ash}
J.~Tian, J.~Hou and B.~Chen,
``Holographic Correlators on Integrable Superstrata,''
Nucl. Phys. B \textbf{948}, 114766 (2019)
[arXiv:1904.04532 [hep-th]].

\bibitem{Bena:2020iyw}
I.~Bena, A.~Houppe and N.~P.~Warner,
``Delaying the Inevitable: Tidal Disruption in Microstate Geometries,''
JHEP \textbf{02}, 103 (2021)
[arXiv:2006.13939 [hep-th]].

\bibitem{Martinec:2020cml}
E.~J.~Martinec and N.~P.~Warner,
``The Harder They Fall, the Bigger They Become: Tidal Trapping of Strings by Microstate Geometries,''
JHEP \textbf{04}, 259 (2021)
[arXiv:2009.07847 [hep-th]].

\bibitem{Ceplak:2021kgl}
N.~Ceplak, S.~Hampton and Y.~Li,
``Toroidal tidal effects in microstate geometries,''
JHEP \textbf{03}, 021 (2022)
[arXiv:2106.03841 [hep-th]].

\bibitem{Mayerson:2020tpn}
D.~R.~Mayerson,
``Fuzzballs and Observations,''
Gen. Rel. Grav. \textbf{52}, no.12, 115 (2020)
[arXiv:2010.09736 [hep-th]].

\bibitem{Bacchini:2021fig}
F.~Bacchini, D.~R.~Mayerson, B.~Ripperda, J.~Davelaar, H.~Olivares, T.~Hertog and B.~Vercnocke,
``Fuzzball Shadows: Emergent Horizons from Microstructure,''
Phys. Rev. Lett. \textbf{127}, no.17, 171601 (2021)
[arXiv:2103.12075 [hep-th]].

\bibitem{Bah:2021jno}
I.~Bah, I.~Bena, P.~Heidmann, Y.~Li and D.~R.~Mayerson,
``Gravitational footprints of black holes and their microstate geometries,''
JHEP \textbf{10}, 138 (2021)
[arXiv:2104.10686 [hep-th]].

\bibitem{Ikeda:2021uvc}
T.~Ikeda, M.~Bianchi, D.~Consoli, A.~Grillo, J.~F.~Morales, P.~Pani and G.~Raposo,
``Black-hole microstate spectroscopy: Ringdown, quasinormal modes, and echoes,''
Phys. Rev. D \textbf{104}, no.6, 066021 (2021)
[arXiv:2103.10960 [gr-qc]].

\bibitem{Shigemori:2019orj}
M.~Shigemori,
``Counting Superstrata,''
JHEP \textbf{10}, 017 (2019)
[arXiv:1907.03878 [hep-th]].

\bibitem{Mayerson:2020acj}
D.~R.~Mayerson and M.~Shigemori,
``Counting D1-D5-P microstates in supergravity,''
SciPost Phys. \textbf{10}, no.1, 018 (2021)
[arXiv:2010.04172 [hep-th]].

\bibitem{David:2002wn}
J.~R.~David, G.~Mandal and S.~R.~Wadia,
``Microscopic formulation of black holes in string theory,''
Phys. Rept. \textbf{369}, 549-686 (2002)
[arXiv:hep-th/0203048 [hep-th]].

\bibitem{Avery:2010qw}
S.~G.~Avery,
``Using the D1D5 CFT to Understand Black Holes,''
[arXiv:1012.0072 [hep-th]].

\bibitem{Ceplak:2022wri}
N.~\v{C}eplak, S.~Hampton and N.~P.~Warner,
``Linearizing the BPS Equations with Vector and Tensor Multiplets,''
[arXiv:2204.07170 [hep-th]].

\bibitem{Martinec:2022okx}
E.~J.~Martinec, S.~Massai and D.~Turton,
``On the BPS sector in AdS\_3/CFT\_2 Holography,''
[arXiv:2211.12476 [hep-th]].

\bibitem{Mathur:2003hj}
S.~D.~Mathur, A.~Saxena and Y.~K.~Srivastava,
``Constructing `hair' for the three charge hole,''
Nucl. Phys. B \textbf{680}, 415-449 (2004)
[arXiv:hep-th/0311092 [hep-th]].

\bibitem{Giusto:2013bda}
S.~Giusto and R.~Russo,
``Superdescendants of the D1D5 CFT and their dual 3-charge geometries,''
JHEP \textbf{03}, 007 (2014)
[arXiv:1311.5536 [hep-th]].

{"status": 404, "message": "PID does not exist."}

\bibitem{Lunin:2001jy}
O.~Lunin and S.~D.~Mathur,
``AdS / CFT duality and the black hole information paradox,''
Nucl. Phys. B \textbf{623}, 342-394 (2002)
[arXiv:hep-th/0109154 [hep-th]].

\bibitem{Lunin:2002iz}
O.~Lunin, J.~M.~Maldacena and L.~Maoz,
``Gravity solutions for the D1-D5 system with angular momentum,''
[arXiv:hep-th/0212210 [hep-th]].

\bibitem{Taylor:2005db}
M.~Taylor,
``General 2 charge geometries,''
JHEP \textbf{03}, 009 (2006)
[arXiv:hep-th/0507223 [hep-th]].

\bibitem{Kanitscheider:2007wq}
I.~Kanitscheider, K.~Skenderis and M.~Taylor,
``Fuzzballs with internal excitations,''
JHEP \textbf{06}, 056 (2007)
[arXiv:0704.0690 [hep-th]].

\bibitem{Maldacena:1999bp}
J.~M.~Maldacena, G.~W.~Moore and A.~Strominger,
``Counting BPS black holes in toroidal Type II string theory,''
[arXiv:hep-th/9903163 [hep-th]].

\bibitem{Maldacena:1997de}
J.~M.~Maldacena, A.~Strominger and E.~Witten,
``Black hole entropy in M theory,''
JHEP \textbf{12}, 002 (1997)
[arXiv:hep-th/9711053 [hep-th]].

\bibitem{Bena:2022fzf}
I.~Bena, N.~\v{C}eplak, S.~D.~Hampton, A.~Houppe, D.~Toulikas and N.~P.~Warner,
``Themelia: the irreducible microstructure of black holes,''
[arXiv:2212.06158 [hep-th]].

\bibitem{Bossard:2019ajg}
G.~Bossard and S.~L\"ust,
``Microstate geometries at a generic point in moduli space,''
Gen. Rel. Grav. \textbf{51}, no.9, 112 (2019)
[arXiv:1905.12012 [hep-th]].

\bibitem{Martinec:2017ztd}
E.~J.~Martinec and S.~Massai,
``String Theory of Supertubes,''
JHEP \textbf{07}, 163 (2018)
[arXiv:1705.10844 [hep-th]].

\bibitem{Martinec:2018nco}
E.~J.~Martinec, S.~Massai and D.~Turton,
``String dynamics in NS5-F1-P geometries,''
JHEP \textbf{09}, 031 (2018)
[arXiv:1803.08505 [hep-th]].

\bibitem{Martinec:2019wzw}
E.~J.~Martinec, S.~Massai and D.~Turton,
``Little Strings, Long Strings, and Fuzzballs,''
JHEP \textbf{11}, 019 (2019)
[arXiv:1906.11473 [hep-th]].

\bibitem{Martinec:2020gkv}
E.~J.~Martinec, S.~Massai and D.~Turton,
``Stringy Structure at the BPS Bound,''
JHEP \textbf{12}, 135 (2020)
[arXiv:2005.12344 [hep-th]].

\bibitem{Bufalini:2021ndn}
D.~Bufalini, S.~Iguri, N.~Kovensky and D.~Turton,
``Black hole microstates from the worldsheet,''
JHEP \textbf{08}, 011 (2021)
[arXiv:2105.02255 [hep-th]].

\bibitem{Bufalini:2022wyp}
D.~Bufalini, S.~Iguri, N.~Kovensky and D.~Turton,
``Worldsheet Correlators in Black Hole Microstates,''
Phys. Rev. Lett. \textbf{129}, no.12, 12 (2022)
[arXiv:2203.13828 [hep-th]].

\bibitem{Bufalini:2022wzu}
D.~Bufalini, S.~Iguri, N.~Kovensky and D.~Turton,
``Worldsheet computation of heavy-light correlators,''
[arXiv:2210.15313 [hep-th]].

\bibitem{Bena:2022wpl}
I.~Bena, S.~D.~Hampton, A.~Houppe, Y.~Li and D.~Toulikas,
``The (amazing) Super-Maze,''
[arXiv:2211.14326 [hep-th]].

\bibitem{deBoer:1998us}
J.~de Boer,
``Large N elliptic genus and AdS / CFT correspondence,''
JHEP \textbf{05}, 017 (1999)
[arXiv:hep-th/9812240 [hep-th]].


\bibitem{Giusto:2013rxa}
S.~Giusto, L.~Martucci, M.~Petrini and R.~Russo,
``6D microstate geometries from 10D structures,''
Nucl. Phys. B \textbf{876}, 509-555 (2013)
[arXiv:1306.1745 [hep-th]].

\end{thebibliography}
\end{document}